\documentstyle[seceq,preprint]{jpsj}
\def\sf{\rm}
\catcode`\@=11
\newcount\@arga
\newcount\@argb
\newcount\@argc
\newcount\@ctmpa
\newcount\@ctmpb
\newcount\@ctmpc
\newcount\@ctmpd
\newcount\@ctmpe
\newdimen\@darg
\newdimen\@bblen
\newif\if@bbllx
\newif\if@bblly
\newif\if@bburx
\newif\if@bbury
\newif\if@height
\newif\if@width
\newif\if@scale
\newif\ifno@bb
\newif\ifepsfdraft
\epsfdraftfalse
\def\@setpsfile#1{
                \typeout{epsf:[#1]}
                \def\@psfile{#1}
}
\def\@setpsheight#1{
                \@heighttrue
                \@darg=#1\relax
                \edef\@psheight{\number\@darg}
}
\def\@setpswidth#1{
                \@widthtrue
                \@darg=#1\relax
                \edef\@pswidth{\number\@darg}
}
\def\@setpsscale#1{
                \@scaletrue
                \def\@pshscale{#1}
                \def\@psvscale{#1}
                \@bblen=#1pt\relax
                \@bblen=1000\@bblen
                \def\@texhscale{\expandafter\remove@dim\the\@bblen}
                \let\@texvscale=\@texhscale
}
\def\@setpshscale#1{
                \@scaletrue
                \def\@pshscale{#1}
                \@bblen=#1pt\relax
                \@bblen=1000\@bblen
                \def\@texhscale{\expandafter\remove@dim\the\@bblen}
}

\def\@setpsvscale#1{
                \@scaletrue
                \def\@psvscale{#1}
                \@bblen=#1pt\relax
                \@bblen=1000\@bblen
                \def\@texvscale{\expandafter\remove@dim\the\@bblen}
}

\def\@setparms#1=#2,{\@nameuse{@setps#1}{#2}}
\def\ps@init@parms{
                \@heightfalse \@widthfalse
                \no@bbfalse
                \def\@psbbllx{}\def\@psbblly{}
                \def\@psbburx{}\def\@psbbury{}
                \def\@psheight{}\def\@pswidth{}
                \def\@pshscale{1}\def\@psvscale{1}
                \def\@texhscale{1000}\def\@texvscale{1000}
                \def\@psfile{}
                \def\@sc{}
}
\def\parse@ps@parms#1{
                \@for\@epsfile:=#1\do
                   {\expandafter\@setparms\@epsfile,}}
\newif\ifnot@eof
\newread\ps@stream
\def\bb@search{
        \openin\ps@stream=\@psfile
        \no@bbtrue
        \not@eoftrue
        \catcode`\%=12\relax
        \ifeof\ps@stream\typeout{epsf: File not found}\fi
        \loop
                \read\ps@stream to \line@in
                \global\toks200=\expandafter{\line@in}\relax
                \ifeof\ps@stream \not@eoffalse \fi
                \@bbtest{\toks200}\relax
                \if@bbmatch\not@eoffalse\expandafter\bb@cull\the\toks200\fi
        \ifnot@eof \repeat
        \catcode`\%=14
}       

\catcode`\%=12
\newif\if@bbmatch
\def\@bbtest#1{\expandafter\@a@\the#1
\long\def\@a@#1
        \ifx\@bbtest#2\@bbmatchfalse\else\@bbmatchtrue\fi}
\def\bb@cull 
        \@ifnextchar\space{\@latexbug}{\bb@extract}}
\def\bb@extract #1 #2 #3 #4 {
        \message{BoundingBox: (#1bp,#2bp)--(#3bp,#4bp)}
        \@darg=#1 bp\edef\@psbbllx{\number\@darg}
        \@darg=#2 bp\edef\@psbblly{\number\@darg}
        \@darg=#3 bp\edef\@psbburx{\number\@darg}
        \@darg=#4 bp\edef\@psbbury{\number\@darg}
        \no@bbfalse
}
\catcode`\%=14

\def\compute@bb{
                \bb@search
                \ifno@bb \typeout{epsf: No BoundingBox}
                \stop
                \else
                \@arga=\@psbburx
                \advance\@arga by -\@psbbllx
                \edef\@bbw{\number\@arga}
                \@arga=\@psbbury
                \advance\@arga by -\@psbblly
                \edef\@bbh{\number\@arga}
                \fi
}
\def\in@hundreds#1#2#3{\@argb=#2 \@argc=#3
                     \@ctmpa=\@argb     
                     \divide\@ctmpa by \@argc
                     \@ctmpb=\@ctmpa
                     \multiply\@ctmpb by \@argc
                     \advance\@argb by -\@ctmpb
                     \multiply\@argb by 10
                     \@ctmpb=\@argb     
                     \divide\@ctmpb by \@argc
                     \@ctmpc=\@ctmpb
                     \multiply\@ctmpc by \@argc
                     \advance\@argb by -\@ctmpc
                     \multiply\@argb by 10
                     \@ctmpc=\@argb     
                     \divide\@ctmpc by \@argc
                     \@arga=#1\@ctmpe=0
                     \@ctmpd=\@arga
                        \multiply\@ctmpd by \@ctmpa
                        \advance\@ctmpe by \@ctmpd
                     \@ctmpd=\@arga
                        \divide\@ctmpd by 10
                        \multiply\@ctmpd by \@ctmpb
                        \advance\@ctmpe by \@ctmpd
                     \@ctmpd=\@arga
                        \divide\@ctmpd by 100
                        \multiply\@ctmpd by \@ctmpc
                        \advance\@ctmpe by \@ctmpd
                     \edef\@result{\number\@ctmpe}
}
\def\compute@wfromh{
                \in@hundreds{\@psheight}{\@bbw}{\@bbh}
                \edef\@pswidth{\@result}
}
\def\compute@hfromw{
                \in@hundreds{\@pswidth}{\@bbh}{\@bbw}
                \edef\@psheight{\@result}
}
\def\compute@handw{
        \if@height 
                \if@width
                \else
                        \compute@wfromh
                \fi
        \else 
                \if@width
                        \compute@hfromw
                \else
                        \if@scale
                                \in@hundreds{\@texvscale}{\@bbh}{1000}
                                \let\@bbh=\@result
                                \in@hundreds{\@texhscale}{\@bbw}{1000}
                                \let\@bbw=\@result
                        \fi
                                \edef\@psheight{\@bbh}
                                \edef\@pswidth{\@bbw}
                \fi
        \fi
}
{\catcode`\p=12\catcode`\t=12
\gdef\remove@dim#1.#2pt{#1}}

\def\compute@sizes{
        \compute@bb
        \compute@handw
}
\def\epsfile#1{
        \ps@init@parms
        \parse@ps@parms{#1}
        \compute@sizes
        \@arga=\@psheight
        \divide\@arga by 65536
        \edef\@psvsize{\number\@arga}
        \@arga=\@pswidth
        \divide\@arga by 65536
        \edef\@pshsize{\number\@arga}
        \message{=>(\@pshsize bp,\@psvsize bp)}
        \leavevmode
        \vbox to \@psheight true sp{
                \hbox to \@pswidth true sp{
                \ifepsfdraft\hss\@psfile\hss\else
                \if@height 
                        \if@width
                                \special{epsfile=\@psfile \space 
                                hsize=\@pshsize \space
                                vsize=\@psvsize \space}
                        \else
                                \special{epsfile=\@psfile \space 
                                vsize=\@psvsize \space}
                        \fi
                \else 
                        \if@width
                                \special{epsfile=\@psfile \space 
                                hsize=\@pshsize \space}
                        \else
                                \if@scale
                                        \special{epsfile=\@psfile \space
                                        vscale=\@psvscale \space
                                        hscale=\@pshscale \space}
                                \else
                                        \special{epsfile=\@psfile \space}
                                \fi
                        \fi
                \fi
                \hfil\fi
                }
        \vfil
        }
}

\catcode`\@=12    
\newcommand{\eps}{\varepsilon}

\newcommand{\vphi}{\varphi}
\def\runtitle{Macroscopic Quantum Tunneling of a Fluxon
in a Long Josephson Junction}
\def\runauthor{Takeo {\sc Kato} and Masatoshi {\sc Imada}}
\title
{Macroscopic Quantum Tunneling of a Fluxon \\
in a Long Josephson Junction}

\author
{Takeo {\sc Kato} and Masatoshi {\sc Imada} }

\inst
{Institute for Solid State Physics, 
University of Tokyo,\\ Roppongi, Minato-ku, Tokyo 106}

\recdate{April 15, 1996}

\abst{
Macroscopic quantum tunneling (MQT) for a single fluxon 
moving along a long Josephson junction is studied theoretically.
To introduce a fluxon-pinning force, we consider inhomogeneities 
made by modifying thickness of an insulating layer locally.
Two different situations are studied: one is the quantum tunneling
from a metastable state caused by a single inhomogeneity, and the other
is the quantum tunneling in a two-state system made by two inhomogeneities.
In the quantum tunneling from a metastable state, 
the decay rate is estimated 
within the WKB approximation. Dissipation effects on a fluxon dynamics are
taken into account by the Caldeira-Leggett theory. 
We propose a device to observe quantum tunneling
of a fluxon experimentally. 
Required experimental resolutions
to observe MQT of a fluxon seem attainable
within the presently available micro-fabrication technique.
For the two-state system, we study
quantum resonance between two stable states, i.e., macroscopic 
quantum coherence (MQC). 
From the estimate for dissipation coefficients due to quasiparticle
tunneling, the observation of MQC appears to be possible within the
Caldeira-Leggett theory.}

\kword
{macroscopic quantum tunneling, fluxon, 
long Josephson junction, dissipation,
Caldeira-Leggett theory, sine-Gordon equation, 
macroscopic quantum coherence} 
\begin{document}
\sloppy
\maketitle
\section{Introduction}
It is commonly recognized that the sine-Gordon equation plays an
outstanding role in many physical problems.
One of the most important applications
is a long Josephson junction (LJJ). 
When the junction width is taken 
large enough in one direction
(defined as $x$-direction), a phase difference $\phi$ of superconductors 
across the junction may have spatial dependence in the $x$-direction. 
It is believed that dynamics of the phase difference $\phi$
is well described by a classical equation~\cite{Pedersen}
\begin{equation}
\phi_{tt} - \phi_{xx} + \sin \phi   
         + \alpha \phi_t - \beta \phi_{xxt} + f = 0 .
\label{SGE}
\end{equation}
Here $x$ and $t$ are measured in units of the Josephson
penetration length $\lambda_{\rm J}$ and of the inverse Josephson
plasma frequency $\omega_{\rm p}^{-1}$, respectively. The dissipation coefficient
$\alpha$ is related to quasiparticle tunneling through
the oxide barrier, and $\beta$ is related to the normal
current of quasiparticle parallel to the junction. 
The external current $f$ is assumed spatially uniform.

It is known that solitons in the form of fluxons
propagate along the junction following the classical equation (\ref{SGE}).
Experimentally, fluxons in LJJs were first observed
indirectly by zero-field steps on the current voltage (I-V) 
characteristics of the junction.~\cite{Fulton}
The zero-field steps are well explained by fluxon propagation
governed by eq.~(\ref{SGE}) with repeated reflections 
at the open ends of the junction.\cite{Lomdahl}
Since then, development of experimental techniques has 
made it possible to directly observe profiles
of separate fluxons and a space-time pattern of their 
interaction.~\cite{Matsuda,Fujimaki} 

These experiments are basically explained by `classical' theories, and
there is no clear experimental indication
so far which needs `quantum' theories to explain its results. 
One might argue that it is natural to expect an essentially
classical motion for a single fluxon of the size of micrometer.
However, we show in this paper that quantum effects can indeed show up
in this extended object.
Recently, Hermon {\it et al.} discussed the quantum dynamics 
of a single fluxon in a long circular Josephson junction.
\cite{Hermon1} In a subsequent paper,\cite{Hermon2} they
discussed a possible fluxon interference experiment.
However in those papers, decoherence effects due to 
couplings to the environment, i.e., 
dissipation has been neglected.
From the study in dissipative 
two-level systems,~\cite{Leggett1} 
it is expected that the quantum effects are 
strongly suppressed by the dissipation 
so that in real experiments, the observation of the quantum effects
proposed by Hermon {\it et al.}
may be more difficult than in the ideal case without 
the dissipation.

In this paper, we propose other experiments to
observe quantum effects of a single fluxon.
The effect which we
deal with here is quantum tunneling of a fluxon. 
Quantum tunneling phenomenon is one of the most typical ones
which cannot be explained by classical theories.
Because a fluxon is a macroscopic object
with a length scale of micrometer,
the quantum tunneling of a fluxon can be recognized as 
the macroscopic quantum tunneling (MQT).
The macroscopic tunneling phenomena have already been
studied in some other systems. 
For example, much attention has been paid on
MQT of the phase of current-biased junctions and the
dissipation on it. MQT in this system has been observed 
experimentally,~\cite{Voss,Washburn}
and it has been claimed that the tunneling rate agrees with the value
predicted by the Caldeira-Leggett theory~\cite{Caldeira}
within a phenomenological treatment of the dissipation.

\begin{figure}[tbp]
\hfil
\epsfile{file=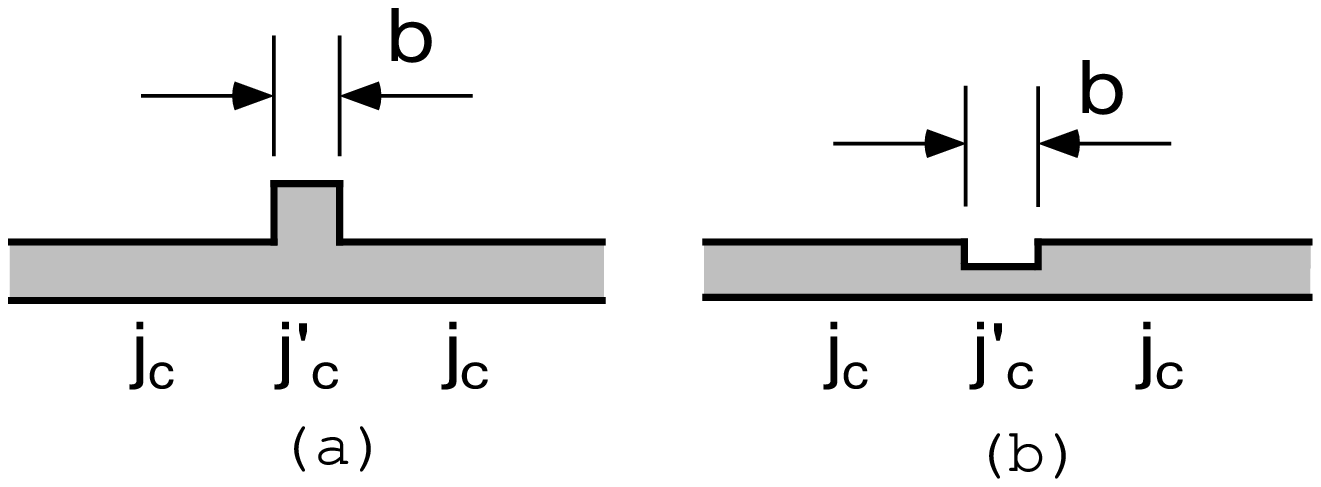,scale=0.8}
\hfil
\caption{Schematic drawings of (a) a microresister 
$(j_{\rm c}>j_{\rm c}')$ and (b) a microshort 
$(j_{\rm c}<j_{\rm c}')$. The normalized strength of pinning
is given by $\eps = b(j_{\rm c}-j_{\rm c}')/j_{\rm c}\lambda_{\rm J}$ 
in both cases. Here $b$ is the length of the area 
where the critical current is modified.}
\label{inhomogenities}
\end{figure}

To consider the quantum tunneling of a fluxon, we introduce
structural inhomogeneities which capture a fluxon at a certain point
in a Josephson junction line.
Such inhomogeneities have been studied
by McLaughlin and Scott in a pioneering paper.~\cite{McLaughlin}
In that paper, they have considered interaction between a fluxon
and a microresister (i.e., a narrow region where critical current density
$j_{\rm c}$ is reduced to $j_{\rm c}'$) or 
a microshort (i.e., an narrow region with a enhanced 
critical current density). Schematic drawings of these inhomogeneities
are shown in Fig.~\ref{inhomogenities}.
McLaughlin and Scott have 
proposed a model of LJJs with a local inhomogeneity
in the form of the equation of motion
\begin{equation}
\phi_{tt} - \phi_{xx} + \sin \phi   
         + \alpha \phi_t - \beta \phi_{xxt} + f 
         - \eps \delta (x) \sin \phi = 0 .
\label{SGEI}
\end{equation}
The last term in the left hand side represents a local change
of the critical current density at $x=0$. The normalized
strength of the inhomogeneity, $\eps$ is defined by
$\eps = (j_{\rm c} - j_{\rm c}')b/j_{\rm c} \lambda_{\rm J}$.
The cases $\eps>0$ and $\eps<0$ correspond to a microresister
and a microshort, respectively.  As we show later,
when $\eps>0$ (i.e. a microresister), the fluxon is attracted 
to the microresister and 
captured there if the fluxon does not have enough kinetic energy. 
In other words, the microresister plays a role of a pinning potential
for the fluxon. We only deal with the case $\eps>0$, i.e., microresisters.
Throughout this paper, we assume $b \ll \lambda_{\rm J}$ 
(i.e. $\eps \ll 1$) so that the soliton
size is larger than $b$ to justify the description by $\delta(x)$ in
(\ref{SGEI}) for the microresister.

In this paper, we study the quantum tunneling of a fluxon
in two situations. First, we consider the situation 
that there exists only one 
microresister in the junction. When the external current $f$
is applied on the junction, the fluxon-pinning state becomes metastable.
We study quantum tunneling of a single fluxon from the metastable state.
The decay rate is estimated within the framework of 
the WKB approximation by taking proper values of experimental parameters.
We also propose experimental design to observe quantum tunneling of a
fluxon, and study conditions to allow the observation in this
device. Next, we study the quantum tunneling 
between two stable states made by two microresisters.
In this situation, we pursue possibility
to observe resonance between two localized levels, i.e., 
macroscopic quantum coherence.
In both cases, we consider dissipation effects  
using the Caldeira-Leggett theory.

The paper is organized as follows.
In $\S$~\ref{faction}, we introduce a model Hamiltonian
for fluxons in the analogy to the Caldeira-Leggett theory, and derive 
the effective action which describes dynamics of a fluxon. 
In $\S$~\ref{parameters}, we summarize controllable experimental parameters,
and set up them to the values which are accessible 
in practical experiments. Some quantities characteristic of LJJs,
are also estimated.  
In $\S$~\ref{metastable}, we consider quantum tunneling
of a single fluxon from a metastable state, and
in $\S$~\ref{twostate}, we study quantum tunneling of a single fluxon 
in a two-state system made by two microresisters.
Concluding remarks are given in $\S$~\ref{discussion}.

\section{Formulation}\label{faction}
\subsection{Classical equation of motion for a single fluxon}
In this subsection, we derive a classical equation of motion for a fluxon.
For this purpose, we analyze the equation (\ref{SGEI}) 
with the assumption that parameters, $\alpha$, $\beta$,
$f$ and $\eps$ are 
all small. In this case, eq.~(\ref{SGEI}) can be considered 
as a perturbed sine-Gordon equation. The soliton solution 
of the unperturbed sine-Gordon equation
takes the form of a kink,
\begin{equation}
        \phi_0(x,t;u) = 4 \arctan \left[ 
                        \, \exp \left( \frac{x-q(t)}{(1-u^2)^{1/2}} 
                        \right) \right] ,
\label{soliton}
\end{equation}  
which corresponds to a fluxon in LJJs.
Here $q(t) = ut$ is the center coordinate of the kink, and $u$ is the
kink velocity ($|u|<1$). The velocity is normalized by the light velocity,
$\bar c = \lambda_{\rm J} \omega_{\rm p}$. In this paper,
we only discuss the nonrelativistic limit $|u| \ll 1$. 

Based on the kink solution (\ref{soliton}), the classical perturbation
theory can be applied to eq.~(\ref{SGEI}).
McLaughlin and Scott
have shown that the perturbations only affect dynamics
of the center coordinate $q(t)$ and do not change the form
of the kink within the framework of the lowest 
approximation.~\cite{McLaughlin}
The equation of motion for $q(t)$ is obtained as
\begin{equation}
        m \ddot q 
        + m \Bigl( \alpha + \frac{\beta}{3} \Bigr) \dot q
        + \frac{\partial V(q)}{\partial q} = 0 ,
\label{EOMF}
\end{equation}
in the nonrelativistic limit. Here $m$ is the classical soliton mass  
and identically equals to 8. Note that the damping strength
working on the dynamics of a fluxon is related to $\alpha$ and $\beta$ 
which are the coefficients of the dissipative terms 
in the field equation (\ref{SGEI}) for $\phi(x,t)$.
The effective potential $V(q)$ for a fluxon is given by
\begin{equation}
        V(q) = - 2\pi f q -\frac{2\eps}{\cosh^2q} .
\end{equation}
The first term is a driving force due to the external current,
and the second term is a pinning potential caused by the microresister
located at the origin. 
%
\subsection{The Caldeira-Leggett theory}\label{CLtheory}
As shown in eq.~(\ref{EOMF}), the classical equation of motion
for a fluxon includes a damping term due to the fact
that the phase difference $\phi$ is a macroscopic variable.
To deal with dissipation effects on MQT of a fluxon phenomenologically, 
the Caldeira-Leggett model~\cite{Caldeira} is introduced
in this subsection.
We provide harmonic oscillators coupled to the macroscopic variable $q$,
and write the Hamiltonian as
\begin{equation}
        H=\frac{p^2}{2m}+V(q) 
         + \sum_{j} \frac12
                \left[ \frac{p_j^2}{m_j} + m_j \omega_j^2
                \biggl( x_j - \frac{c_j}{m_j \omega_j^2} q \biggr)^2
                \, \right] . 
\label{CLHamiltonian}
\end{equation} 
Here $p=m\dot q$ is a momentum variable conjugate to $q$, whereas
$q_j$ and $p_j=m_j\dot q_j$ are a coordinate and a momentum
of harmonic oscillators, respectively. 
The Caldeira-Leggett model has succeeded
in explaining experiments of junctions phenomenologically.~\cite{Washburn}
Here, we extend this treatment to LJJs.
The reservoir parameters, $m_j$, $\omega_j$
and $c_j$ are characterized by a spectral function
\begin{equation}
        J(\omega) = \frac{\pi}{2} 
                \sum_{j} \frac{c_j^2}{m_j\omega_j}
                \delta(\omega-\omega_j) .
        \label{defspectral}
\end{equation}
We choose the spectral function as
\begin{equation}
        J(\omega)=m\Bigl(\alpha+\frac{\beta}{3}\Bigr)\omega.
\end{equation}
Then, the classical equation (\ref{EOMF})
is derived from the Hamiltonian (\ref{CLHamiltonian})
by eliminating the reservoir degrees of freedom.~\cite{Caldeira} 

From this model, we obtain the partition function $Z$ 
after integrating out the reservoir degrees of freedom as
\begin{equation}
        Z = Z_{\rm R}\oint {\cal D} q(\tau)
                \exp \Bigl( - \frac{1}{g^2}
                S_{{\rm eff}}\,[\,q(\tau)\,] \Bigr).
\end{equation}
Here $Z_{\rm R}$ is the partition function of the unperturbed reservoir, and
$g^2$ is the normalized Planck constant defined by 
$g^2=\hbar\omega_{\rm p}/E_0$
where $E_0$ is the energy scale in the LJJ. 
The effective action $S_{{\rm eff}}$ is calculated as~\cite{Weiss}
\begin{equation}
        S_{{\rm eff}}[q(\tau)] 
                = \int_0^{1/T} {\rm d}\tau \biggl(
                        \frac12 m\dot q^2 + V(q) \biggr) 
                + \frac12 \int_0^{1/T} {\rm d}\tau
                \int_0^{\tau} {\rm d}\tau' K(\tau-\tau')
                \Bigl( q(\tau)-q(\tau') \Bigr)^2 ,
        \label{CaldeiraSeff} 
\end{equation}
where $T$ is a temperature normalized by $\hbar \omega_{\rm p}/k_{\rm B}$.
Here, $k_{\rm B}$ is Boltzmann's constant.
The kernel $K(\tau)$ is given by
\begin{eqnarray}
        K(\tau)&=&\frac{1}{\pi}\int_0^{\infty} 
                m\Bigl(\alpha+\frac{\beta}{3}\Bigr)\omega D_{\omega}(\tau) 
        \, {\rm d}\omega, \\
        D_{\omega}(\tau) 
                &=& \frac{\cosh \bigl[\:\omega(1/2T
                - |\tau| \, ) \: \bigr] }{\sinh(\omega/2T)} .
        \label{DWDEF}
\end{eqnarray}
In the later sections, we make estimate of the tunneling rate
based on the effective action (\ref{CaldeiraSeff}).

The derivation of the effective action (\ref{CaldeiraSeff})
is, however, quite intuitive. In the above derivation,
infinite degrees of freedom of the
field $\phi$ is first reduced to only one degree of freedom, i.e.,
the center coordinate $q$ in the equation of motion
(\ref{EOMF}). Then, quantization for $q$ is performed. If
the quantum fluctuation around a fluxon affects
quantum tunneling, the effective action (\ref{CaldeiraSeff})
may take a different form, because the fluctuation of the field 
is eliminated in the derivation of (\ref{EOMF}) before the quantization.
In the next subsection, we check the validity
of the effective action (\ref{CaldeiraSeff}) 
by applying the Caldeira-Leggett formalism directly to the equation
of motion for the field variable, eq.~(\ref{SGEI}).
The readers who are not interested in details of theoretical derivation
may skip the next subsection.

\subsection{Another derivation of the effective action}
\label{semicl}
In this subsection, the effective action (\ref{CaldeiraSeff})
is derived in another way. The plan is as follows.
First, quantization for the field variable $\phi$ is
performed. The dissipation effects are taken into account 
by constructing a model Hamiltonian on the analogy of the
Caldeira-Leggett theory. 
Then, the semiclassical theory is applied to the model
with the path integral method.
Through the perturbation expansion
for the parameters, $\alpha$, $\beta$, $f$ and $\eps$, 
we show that the effective action (\ref{CaldeiraSeff}) is reproduced
within the lowest order approximation.

First, we derive the Hamiltonian which yields classical 
field equation (\ref{SGEI}). In the absence of dissipation, 
the Hamiltonian is easily obtained as
\begin{equation}
        H_{\rm S} = \int {\rm d}x \Bigl[ \frac12 \phi_x^2 + \frac12 \phi_t^2 
                + ( 1 - \cos \phi ) + f \phi - 
        \eps \delta(x) (1-\cos \phi) \Bigr] .
        \label{Hs}
\end{equation}
In the dissipative case, the dissipative terms,
$-\alpha \phi_{t}$ and $\beta \phi_{xxt}$ are obtained
by introducing two kinds of harmonic
oscillators coupled linearly via the field variables $\phi (x,t)$ and
$\phi_{x} (x,t)$ as follows:
\begin{eqnarray}
        H_{\alpha} &=& \int {\rm d}x \sum_{j} \Biggl[ \: 
                \frac{p_j^2(x)}{2m_j} 
                + \frac{m_j\omega_j^2}{2}
                \left( q_j(x) 
        - \frac{c_j}{m_j\omega_j^2} \phi(x) 
        \right)^2 \, \Biggr] , \label{Ha}         \\
        H_{\beta} &=& \int {\rm d}x \sum_j \Biggl[ \:
                \frac{p_j'^2(x)}{2m_j'} 
                + \frac{m_j'\omega_j'^2}{2}
                \left( q_j'(x) 
        - \frac{c_j'}{m_j'\omega_j'^2} \phi_x(x) 
        \right)^2 \, \Biggr] . \label{Hb} 
\end{eqnarray}
These oscillators constitute a proper heat bath causing the dissipation.

The reservoir is characterized by the spectral functions
$J_{\alpha}(\omega)$ and $J_{\beta}(\omega)$ defined by
\begin{eqnarray}
        J_{\alpha}(\omega) &=& \frac{\pi}{2} \sum_j
                 \frac{c_j^2}{m_j\omega_j}
                        \delta(\omega - \omega_j) , 
        \label{ja} \\
        J_{\beta}(\omega) &=& \frac{\pi}{2} \sum_j
                 \frac{c_j'^2}{m_j'\omega_j'}
                        \delta(\omega - \omega_j') . 
        \label{jb}
\end{eqnarray}
In order to produce the dissipative terms, $-\alpha \phi_t$ 
and $\beta \phi_{xxt}$, we choose the spectral functions as
\begin{eqnarray}
        J_{\alpha}(\omega) &=& \alpha \omega , \\
        J_{\beta }(\omega) &=& \beta  \omega .  
\end{eqnarray}
The total Hamiltonian is given by
\begin{equation}
        H = H_{\rm S} + H_{\alpha} + H_{\beta} .
        \label{fieldHtot}
\end{equation}
From this Hamiltonian, the perturbed sine-Gordon equation
(\ref{SGEI}) is derived in the classical limit. 
The details of the derivation is given in Appendix~\ref{sfapp1}.
The dissipative sine-Gordon system
with the Hamiltonian $H_{\alpha}$ has been used
in the study of charge density wave~\cite{Hida} and long Josephson
junctions.~\cite{Simanjuntak} It should be noted that the dissipation 
term $\beta \phi_{xxt}$ in (\ref{SGEI})
is expressed in a simple way by the coupling to the derivative 
$\phi_{x}$. 

After integrating out the the reservoir degrees of freedom,
we obtain the partition function $Z$ as
\begin{equation}
        Z=Z_{\rm R}^{(\alpha)} Z_{\rm R}^{(\beta)} \oint {\cal D}\phi
                \exp\left( -\frac{1}{g^2}
                S_{{\rm eff}}[\phi(x,\tau)] \right).
        \label{FieldActionS}
\end{equation}
Here $Z_{\rm R}^{(\alpha)}$ and $Z_{\rm R}^{(\beta)}$ are partition 
functions of the unperturbed harmonic oscillations, and
$g^2$ is the normalized Plank constant defined 
in $\S$~\ref{CLtheory}. The effective action 
$S_{{\rm eff}}$ is given by 
\begin{equation}
        S_{{\rm eff}} = S_{\rm S} + S_{\alpha} + S_{\beta} ,
\end{equation}
where
\begin{full}
\begin{eqnarray}
        S_{\rm S} &=& \int_0^{1/T} {\rm d}\tau \int {\rm d}x
                 \left( \frac12\phi_x^2+\frac12\phi_t^2
                +(1-\cos\phi) +f\phi - \eps \delta(x)
                \:(1-\cos\phi) \right) , 
        \label{fieldSs}\\
        S_{\alpha} &=& \frac12 \int_0^{1/T} {\rm d}\tau
                \int_0^{\tau} {\rm d}\tau' \int {\rm d}x 
                K_{\alpha}(\tau-\tau') 
                \Bigl( \phi(x,\tau)-\phi(x,\tau') \Bigr)^2 ,
        \label{fieldSa}\\ 
        S_{\beta} &=& \frac12 \int_0^{1/T} {\rm d}\tau 
                \int_0^{\tau} {\rm d}\tau' \int {\rm d}x
                K_{\beta}(\tau-\tau')
                \Bigl( \phi_x(x,\tau)-\phi_x(x,\tau') \Bigr)^2 .
        \label{fieldSb} 
\end{eqnarray}
\end{full}
The kernels, $K_{\alpha}(\tau)$ and $K_{\beta}(\tau)$ are given by
\begin{eqnarray}
        K_{\alpha}(\tau) &=& \frac{1}{\pi} \int_0^{\infty} \!
                \alpha \omega D_{\omega}(\tau) {\rm d}\omega , 
        \label{Ka} \\
        K_{\beta}(\tau) &=& \frac{1}{\pi} \int_0^{\infty} \!
                \beta \omega D_{\omega}(\tau) {\rm d}\omega ,
        \label{Kb}
\end{eqnarray}
where $D_{\omega}(\tau)$ is defined by (\ref{DWDEF}).

Here we derive an effective theory of a single fluxon
by approximating the quantization of the field theory 
defined by (\ref{fieldSs})-(\ref{fieldSb}).
For this purpose, we consider the semiclassical theory for the
sine-Gordon equation. The semiclassical quantization of the
field theory has been studied extensively 
in the context of high-energy physics in 1970's.~\cite{Jackiw,Rajaraman} 
In the semiclassical method, the classical soliton solution is
regarded as the ground state of a Fock space called the one-soliton sector.
This sector is completely separated from the sector containing no soliton,
because of the topological stability of the soliton.
The states of the one-soliton sector are constructed by a perturbative
expansion in $g^2$. This expansion is valid 
as long as $g^2$ is small. As shown in $\S$~\ref{parameters},
$g^2$ is estimated as $g^2 < 10^{-2}$ in proper choices of 
experimental parameters. Hence, the semiclassical approach can be applied.
In this subsection, we only explain the outline of this
approach. The details of the semiclassical calculation is given 
in Appendix~\ref{sfapp2}.

First we consider the unperturbed sine-Gordon theory. The partition
function is given by
\begin{equation}
        Z_0 = \oint {\cal D} \phi(x,\tau) \exp \left(
                -\frac{1}{g^2} S_0[\,\phi(x,\tau)\,] \right) .
        \label{Z0defapp}
\end{equation}
Here $S_0$ is the Euclidean action,
\begin{equation}
        S_0 = \int {\rm d}\tau \int {\rm d}x \left(\frac12 \dot{\phi}^2
                +\frac12 \phi_x^2 +(1-\cos \phi ) \right) ,
        \label{S0defapp}
\end{equation}
where $\displaystyle{\dot{\phi} = \frac{\partial \phi}{\partial t}}$.
This field theory possesses a nontrivial stationary solution
\begin{equation}
        \phi_0(x-q) = 4 \arctan \Bigl[ \, \exp (x-q) \, \Bigr] ,
        \label{SolitonSolution}
\end{equation}
where $q$ is the center coordinate of the soliton. 
This new variable $q$ is called a collective coordinate in the field theory,
and regarded as a dynamical variable. In the semiclassical approach,
we only consider the paths around the stationary solution 
because they predominantly contribute to the partition
function (\ref{Z0defapp}).  The deviation around the
stationary solution is denoted by $\eta$ as
\begin{equation}
        \phi(x,\tau) = \phi_0(x-q(\tau)) + g\eta(x-q(\tau),\tau) .
\end{equation}
In terms of LJJs,
the collective coordinate $q(\tau)$ describes dynamics of a single fluxon,
and the rest degrees of freedom, $\eta(x,\tau)$ represent the 
fluctuation around the fluxon. The perturbation
expansion of the action (\ref{S0defapp}) in terms of $\eta$ is 
nothing but that in terms of $g^2$. In the lowest approximation,
we obtain `plasmon' excitations by the quantization of the 
quadratic parts for $\eta$ in the action.
The plasmon excitations have a energy gap of $\hbar\omega_{\rm p}$. 
Since the energy gap is estimated as several Kelvin,
the plasmon excitation can be neglected at low enough temperatures
of the order of mK.
The higher order terms of $\eta$ in the action 
only contribute to the partition function as the higher order of $g^2$.
The semiclassical expansion of $g^2$ needs careful 
treatments especially in the treatment of the collective
coordinate $q$.~\cite{Jackiw} 
(See Appendix~\ref{sfapp2}.) We only write the result:
\begin{equation}
        Z_0 \approx 
        \oint {\cal D}q(\tau) \exp\left(-\frac{1}{g^2}\int {\rm d}\tau
        \Bigl( m +\frac m2 \dot q^2 \Bigr) \right) .
\end{equation}
From this result, we find that the fluxon behaves like a free particle
with the mass $m=8$. 
It is a natural result because there is no force which
modifies the velocity of the soliton.

Next we consider the perturbed sine-Gordon equation (\ref{SGE}). 
The perturbation part of the action is given by
\begin{eqnarray}
        S_{{\rm ext}}[\phi(x,\tau)]&=&\int_0^{1/T} 
        \! \! {\rm d}\tau \int \! {\rm d}x 
        \, f \, \phi , 
        \label{PAction1}        \\
        S_{{\rm pin}}[\phi(x,\tau)]&=&\int_0^{1/T} 
        \! \! {\rm d}\tau \int \! {\rm d}x 
        \eps \delta(x) (1-\cos \phi) , 
        \label{PAction2}        
\end{eqnarray}
in addition to (\ref{fieldSa}) and (\ref{fieldSb}).
Here we assume that the perturbation parameters, $\alpha$, $\beta$, 
$f$, $\eps$ are all small. Then, we take the perturbation 
expansion in terms of these small terms 
in addition to the expansion in $g^2$.
From straightforward calculation,~\cite{Gervais} 
it is shown that the lowest order contribution of the perturbation
expansion is
obtained only by substituting the soliton solution 
(\ref{SolitonSolution}) to the perturbative
actions. In other words, the perturbations
do not modify the waveform of the soliton in the lowest order.

First, we consider the external current term (\ref{PAction1}).
Substitution of (\ref{SolitonSolution}) gives
\begin{eqnarray}
        S_{{\rm ext}}'[q(\tau)] &=& S_{{\rm ext}}[\phi_0(x-q(\tau))] 
                - S_{{\rm ext}}[\phi_0(x)] \nonumber \\
        &=& \int_0^{1/T} {\rm d}\tau ( -2\pi f q ) .
\end{eqnarray}
Here the prime implies the effective action for the variable $q(\tau)$.
We subtracted the constant $S_{{\rm ext}}[\phi_0(x)]$, because 
we have chosen the origin of the potential energy at $q=0$.

The pinning potential is obtained by substituting (\ref{SolitonSolution})
to (\ref{PAction2}),
\begin{eqnarray}
        S_{{\rm pin}}'[q(\tau)] &=& S_{{\rm pin}}[\phi_0(x-q(\tau))] 
        \nonumber \\
        &=& \int_0^{1/T} {\rm d}\tau \frac{-2\eps}{\cosh^2 q} .
\end{eqnarray}
\begin{full}
Finally, we consider the dissipation described by (\ref{fieldSb})
and (\ref{fieldSa}). The effective action is obtained as
\begin{eqnarray} 
        S_{\alpha}'[q(\tau)] &=& S_{\alpha}[\phi_0(x-q(\tau))] 
        \nonumber \\
        &=& \frac12 \int_0^{1/T} {\rm d}\tau 
        \int_0^{\tau} {\rm d}\tau' \int {\rm d}x K_{\alpha}(\tau-\tau')
                \Bigl(\phi_0\bigl(x-q(\tau)\bigr)
                -\phi_0\bigl(x-q(\tau')\bigr)\Bigr)^2 ,
        \label{Dissa}\\
        S_{\beta}'[q(\tau)] &=& S_{\beta}[\phi_0(x-q(\tau))] 
        \nonumber \\
        &=& \frac12 \int_0^{1/T} {\rm d}\tau 
        \int_0^{\tau} {\rm d}\tau' \int {\rm d}x K_{\beta}(\tau-\tau')
                \Bigl(\phi_0'\bigl(x-q(\tau)\bigr)
                -\phi_0'\bigl(x-q(\tau')\bigr)\Bigr)^2 ,
        \label{Dissb}
\end{eqnarray}
where $\phi_0' 
= \displaystyle{\frac{\partial \phi_0}{\partial x}}$.
Since the integration over $x$ is difficult to carry analytically,
we assume that the paths of the collective coordinate
$q(\tau)$ satisfy
\begin{equation}
        \left| q(\tau)-q(\tau')\right| \ll 1
        \label{condition}
\end{equation}
for all $\tau$, $\tau'$. This condition is well satisfied 
in the case we consider in this paper. Using the approximations
\begin{eqnarray}
        \phi_0(x-q(\tau))-\phi_0(x-q(\tau')) 
        &\simeq& - \phi_0'(x-q(\tau)) \, \Bigl( q(\tau)-q(\tau')\Bigr) , \\
        \phi_0'(x-q(\tau))-\phi_0'(x-q(\tau')) 
        &\simeq& -\phi_0''(x-q(\tau)) \, \Bigl( q(\tau)-q(\tau')\Bigr) ,
\end{eqnarray}
and the identities
\begin{equation}
        \int {\rm d}x \phi_0'^2 = 8 \equiv m \quad \mbox{and} \quad
        \int {\rm d}x \phi_0''^2 = \frac m3 ,
\end{equation}
we obtain from (\ref{Dissa}) and (\ref{Dissb})
\begin{eqnarray}
        S_{\alpha}'[q(\tau)] &=& \frac12 \int_0^{1/T} {\rm d}\tau 
                \int_0^{\tau} {\rm d}\tau'
                m K_{\alpha}(\tau-\tau') 
                \Bigl(q(\tau)-q(\tau')\Bigr)^2 \\
        S_{\beta}'[q(\tau)] &=& \frac12 \int_0^{1/T} {\rm d}\tau 
                \int_0^{\tau} {\rm d}\tau'
                \frac{m}{3} K_{\beta}(\tau-\tau') 
                \Bigl(q(\tau)-q(\tau')\Bigr)^2 .
\end{eqnarray}
\end{full}

We summarize the total effective action of a fluxon.
The partition function is given by
\begin{equation}
        Z = \oint {\cal D}q(\tau) 
        \exp\left( -\frac{1}{g^2} S_{{\rm eff}}'[q(\tau)]\right) 
\end{equation}
where
\begin{eqnarray}
        S'_{{\rm eff}}[q(\tau)]
        &=& \int {\rm d}\tau \left( \frac12 m\dot q^2 + V(q) \right) 
        + \frac12 \int {\rm d}\tau \int {\rm d}\tau' K(\tau-\tau')
                \Bigl(q(\tau)-q'(\tau)\Bigr)^2 \label{SimpleAction} 
        \\
        V(q) &=& -2\pi f q - \frac{2\eps}{\cosh^2(q)}
        \label{generalpotential}
\end{eqnarray}
The kernel $K(\tau)$ is given by
\begin{eqnarray}
        K(\tau) &=& \frac{1}{\pi} \int m\Bigl( \alpha+
                \frac{\beta}{3} \Bigr) D_{\omega}(\tau) {\rm d}\omega,
        \label{Kernel2}
\end{eqnarray}
where $D_{\omega}(\tau)$ is defined in eq.~(\ref{DWDEF}).
The effective action obtained here agrees with 
one given in eq.~(\ref{CaldeiraSeff}), and it has been shown
that more intuitive derivation in $\S$~\ref{CLtheory} is justified. 
One might feel that this result is trivial. It is, however, nontrivial 
because the fluxon has a finite size
in space. In fact, if the condition (\ref{condition}) does not
hold, the action has a different form 
from the simple action (\ref{SimpleAction}).

\section{Experimental Parameters}\label{parameters}
\begin{figure}[tbp]
\hfil
\epsfile{file=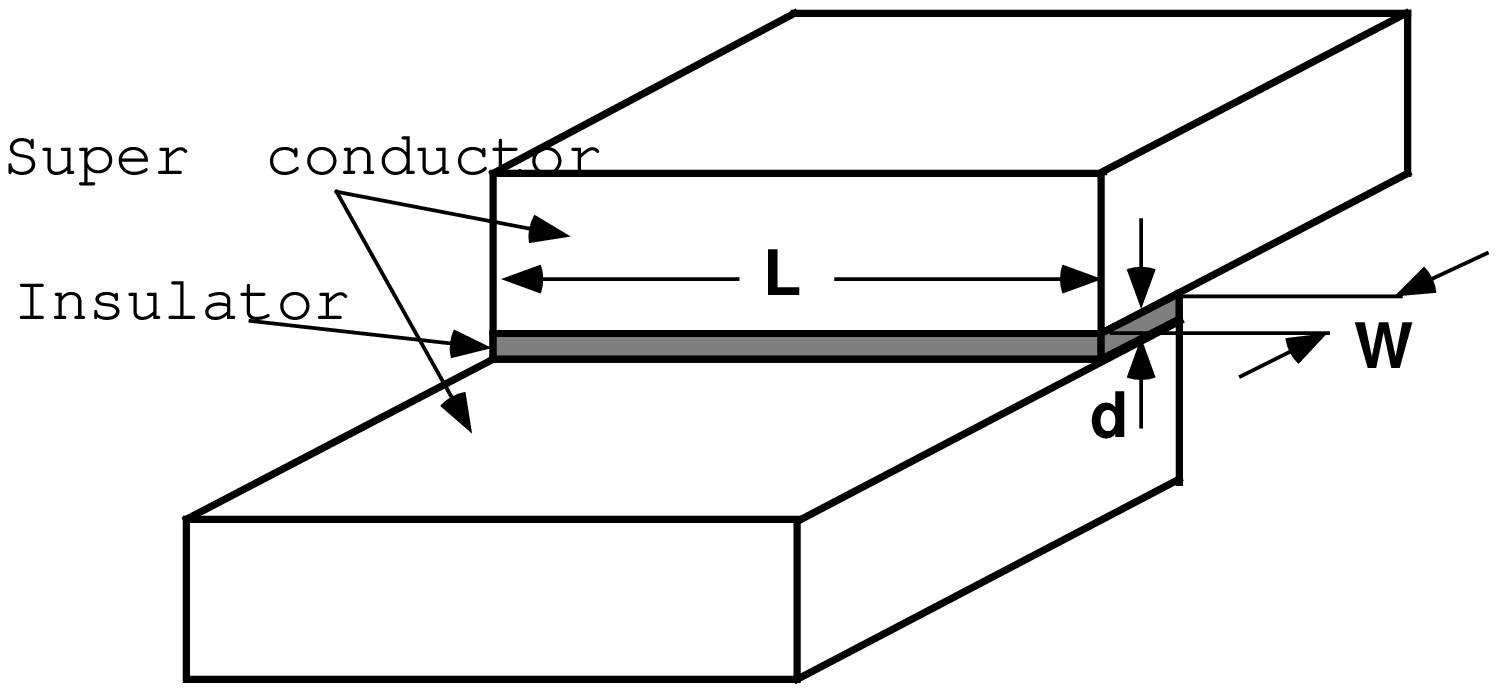,scale=0.65}
\hfil
\caption{A layout of a LJJ. The junction length $L$ is taken as
large enough compared with the junction width $W$.}
\label{Fig1z}
\end{figure}

In this section, we summarize controllable experimental parameters
relevant to the quantum tunneling of a fluxon.
A layout of a LJJ is shown in Fig.~\ref{Fig1z},
and experimentally parameters realizable within the available technique
are given in Table~\ref{expparameter}.
The values given in the table are indeed typical ones in actual
experiments.~\cite{Davidson}
The width of the Josephson junction $W$ is now a controllable 
variable and scaled
by micrometer.
\begin{table}[tbp]
\caption{Experimental parameters used for an estimate.}
\label{expparameter}
\begin{tabular}{@{\hspace{\tabcolsep}\extracolsep{\fill}}ll} \hline
       Parameters & Estimated value \\
       \hline
       \hline
       $d$ : Thickness of the junction  & $2$ [nm] \\
       $\eps_{\rm r}$ : Relative dielectric constant   & $8$ \\
       $j_{\rm c}$ : Critical current density 
       & $3\times10^6$ [A/$\mbox{m}^2$] \\
       $W$ : Width of the junction      & W [$\mu$m] (variable) \\
       $\lambda_L$ : London penetration length & 60 [nm] \\
       \hline
\end{tabular}
\end{table}

First, we associate the experimental parameters with the parameters
in the normalized
sine-Gordon equation (\ref{SGE}). In the derivation of (\ref{SGE}),
the normalized length, time and energy are introduced.
In the original unit, the equation of motion without
inhomogeneities is written as
\begin{equation}
        \omega_{\rm p}^{-2} \phi_{tt} - \lambda_{\rm J}^2 \phi_{xx}
        +\sin \phi + \alpha \omega_{\rm p}^{-1} \phi_t 
        - \beta \omega_{\rm p}^{-1} \lambda_{\rm J}^2 \phi_{xxt} + j/j_{\rm c} = 0 .
\end{equation}
Here, $j$ is the external current density applied to the junction. 

The length scale in the LJJ is given by the Josephson penetration 
length $\lambda_{\rm J}$, which may be varied in the range from $20\,[\mu\mbox{m}]$
to $200\,[\mu\mbox{m}]$ by controlling the critical current density 
$j_{\rm c}$.~\cite{Matsuda}
In the present estimate, $\lambda_{\rm J}$ is obtained as
\begin{equation}
        \lambda_{\rm J} = \left( 
                \frac{\Phi_0}{4\pi\mu_0 \lambda_{\rm L}j_{\rm c}}
                \right)^{1/2}
                \sim 27 \: [\mu\mbox{m}] .
        \label{jc1}
\end{equation}
Here $\Phi_0 = h/2e$ 
is the unit flux and $\mu_0 ( = 4\pi \times 10^{-7} 
\: [\mbox{Hm}^{-1}])$ 
is the permeability of the vacuum. 
The size of a fluxon is characterized
by $\lambda_{\rm J}$ and the junction length $L$ must be taken larger
than $\lambda_{\rm J}$ so that at least a single fluxon exists in the junction.  

The time is scaled by $\omega_{\rm p}^{-1}$ 
where $\omega_{\rm p}$ is the plasma frequency,
which is estimated from
the value given in Table~$\ref{expparameter}$ as
\begin{equation}
        \omega_{\rm p} = \left( \frac{2\pi j_{\rm c}d}{\Phi_0\eps_{\rm r}\eps_0}
                \right)^{1/2} \sim 5.1\times 10^{11}\: \mbox{[1/s]} ,
        \label{wpdef}
\end{equation}
where $\eps_0 (=8.85\times10^{-12} \: [\mbox{Fm}^{-1}])$ 
is the dielectric constant of the vacuum. Note that $\omega_{\rm p}$
is independent of $W$. From this estimate,
the plasmon excitation gap is estimated as
\begin{equation}
        \hbar\omega_{\rm p} \sim 3.9 \:\mbox{[K]} .
\end{equation}
Well below this temperature, the plasmon excitations can be neglected.

The energy is measured by a unit energy $E_0$,
\begin{equation}
        E_0 = \frac{\Phi_0}{2\pi} j_{\rm c}W\lambda_{\rm J} 
                \sim 1.93\times10^3 W \: \mbox{[K]} .
        \label{jc2}
\end{equation}
From this, the energy of a single fluxon is obtained,
\begin{equation}
        8E_0 \sim 1.55\times 10^4W \:\mbox{[K]}
\end{equation}
where the factor $8$ comes from the dimensionless mass of
a fluxon. The energy of a single fluxon is so large that  
nucleation of soliton-antisoliton pairs can be neglected
at sufficiently low temperatures.

The normalized Planck constant, $g^2$ is given as
\begin{equation}
        g^2 = \frac{\hbar\omega_{\rm p}}{E_0} 
        = \frac{16\pi}{137}
        \left(\frac{2\lambda_{\rm L}d}{W^2 \eps_{\rm r}}\right)^{1/2} .
\end{equation}
Note that $g^2$ is independent of the critical current $j_{\rm c}$.
In the present estimate, the value of $g^2$ becomes
\begin{equation}
        g^2 \sim \frac{2.0\times 10^{-3}}{W} .
        \label{g2value}
\end{equation}
When $W\sim1 [\mu\mbox{m}]$, we get small value,
$g^2\sim0.002$. Hence, the semiclassical approach 
explained in $\S$~\ref{semicl} is applicable.

The mass of a fluxon is estimated as
\begin{equation}
        m_{{\rm f}} 
        = \frac{8E_0}{\bar c^2} \sim 1.25\times10^{-3} W \: m_e ,
\end{equation}
where $m_e$ is the electron mass and $\bar c$ is defined by
$\bar c = \lambda_{\rm J} \omega_{\rm p}$. Since we find that the fluxon mass 
is remarkably small ($\sim 10^{-3}m_e$ if we take $W$ as $1$ [$\mu$m]),
we can expect substantial quantum effects of fluxons, in spite of 
the large fluxon size and the small normalized Planck
constant. Moreover, it should be noted that $m_{{\rm f}}$ is
proportional to the junction width $W$. Hence, the fluxon mass
can be controlled by changing the value of $W$. 

Finally, we estimate the dissipation coefficients, $\alpha$ and $\beta$.
The coefficient $\alpha$ is related to the quasiparticle resistance
per area $r_{\rm qp}$ as $\alpha = 1/r_{\rm qp} C' \omega_{\rm p}$. 
Here, $C'=\eps_{\rm r}\eps_0/d$ is the capacitance per unit
area. This relation has been obtained experimentally by
Pedersen and Welner.~\cite{Pedersen2} The quasiparticle 
resistance $r_{\rm qp}$ obtained below the gap voltage 
is strongly enhanced at low temperatures.
Hence, the dissipation coefficient $\alpha$ becomes very small at
sufficiently low temperatures of mK order.

To make a comparison, we introduce a dissipation coefficient
$\alpha_{\rm n}$ defined by $\alpha_{\rm n} = 1/r_{\rm n} C' \omega_{\rm p}$.
Here, $r_{\rm n}$ is the normal resistance obtained 
above the gap voltage.
Since $r_{\rm n}$ is much smaller than $r_{\rm qp}$, the dissipation
coefficient $\alpha_{\rm n}$ gives the upper limit for $\alpha$,
i.e., $\alpha \ll \alpha_{\rm n}$.
Then, $r_{\rm n}$ can be associated with $j_{\rm c}$ as~\cite{Ambegaokar}
\begin{equation}
        j_{\rm c} r_{\rm n} = \frac{\pi\Delta_0}{2e} ,
        \label{ABeq}
\end{equation}
at sufficiently low temperatures $(k_{{\rm B}}T \ll \Delta_0)$.
Here $\Delta_0$ is the gap of the superconducter at zero temperature.
From (\ref{wpdef}) and (\ref{ABeq}), we obtain
\begin{equation}
        \alpha_{\rm n} = \frac{\hbar\omega_{\rm p}}{\pi\Delta_0} ,
        \label{aest}
\end{equation}
which gives a good estimate for $\alpha_{\rm n}$.
Using $\Delta_0 = 14 [{\rm K}]$ and (\ref{wpdef}), 
we have $\alpha_{\rm n} = 0.088$. Thus, we evaluate $\alpha \ll
0.088$. 

The dissipation coefficient $\beta$ also originates from quasiparticle
current. It has been observed that the ratio between 
$\alpha$ and $\beta$ is independent of the 
temperature.~\cite{Davidson,Davidson2} Hence, it is expected that
$\beta$ is also strongly suppressed at sufficiently low temperatures.
We estimate the upper bound for $\beta$ as $\beta \ll 0.01$, which is an
experimentally determined value by 
Davidson {\it et al.}~\cite{Davidson,Davidson2} 
at relatively high temperatures ($\sim 4 \, [{\rm K}]$).
This evaluation for $\beta$ is rough, because the experimental
situation in Ref.~\citen{Davidson} is rather different from
ones employed in this paper. 
The dissipation amplitude $\alpha + \beta/3$ is, however, not
changed drastically even if $\beta$ is enlarged several times,
because $\alpha$ is expected to be dominant in the dissipation
amplitude.

On the basis of the values given in this section,
we estimate the tunneling rate in the later sections.
\section{Tunneling from a Metastable State}\label{metastable}
\subsection{Tunneling rate}
\begin{figure}[tbp]
\hfil
\epsfile{file=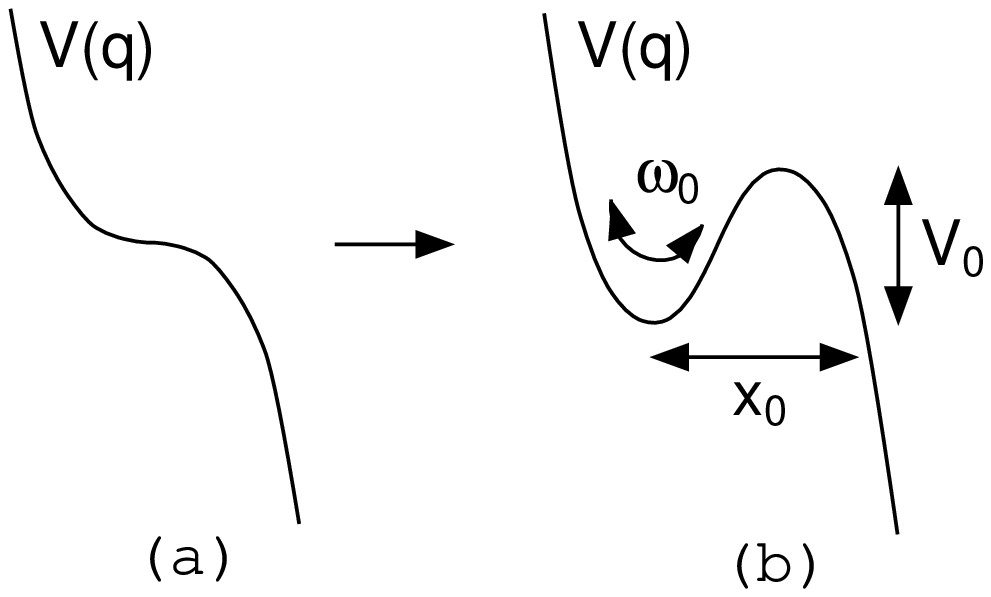,scale=0.8}
\hfil
\caption{The potential form $V(q)$ for a fluxon (a) when the external current
$f$ is taken as the critical value $f_{\rm c}$, and (b) when $f$ is taken
as a slightly smaller value $f=f_{\rm c}-\eta (\eta > 0)$ than $f_{\rm c}$.}
\label{Potential}
\end{figure}

In this section, we consider the quantum tunneling
from a metastable state made by a single microresister in a LJJ.
First, we formulate the decay rate using the Langer's WKB
method.~\cite{Langer,Callan} 
The potential form $V(q)$ is given in (\ref{generalpotential}) as
\begin{equation}
        V(q) = -2\pi f q -\frac{2\eps}{\cosh^2q} .
\end{equation}
The first term is the driving force due to the external current,
and the second term is the pinning potential 
caused by a single microresister.  
If $f$ is small, the potential $V(q)$ has a metastable state at $q=q_0$
defined by $V'(q_0)=0$. However, if $f$ is increased and takes a critical
value $f_{c}$, the metastable state disappears. 
The critical value $f_{c}$ is calculated as
\begin{equation}
        f_{\rm c} = \frac{4\eps}{3\sqrt{3}\pi} \simeq 0.245\eps (\ll 1) .
\label{fcdef}
\end{equation}
When the external current $f$ is taken as $f=f_{\rm c}-\eta$ ($\eta>0$),
the potential energy has a barrier $V_0$ as shown in Fig.~\ref{Potential}.
To observe quantum tunneling in the laboratory,
$V_0$ must be small. Hence, we assume $\eta \ll f_{\rm c}$.
The potential is approximated by the quadratic-plus-cubic 
potential around a metastable state as
\begin{equation}
        V(x) = -\frac{16\eps}{9\sqrt{3}}x^3 
                + \sqrt{\frac{32\pi\eps\eta}{3\sqrt{3}}} x^2 ,
        \label{approxpot}
\end{equation}
where the origin of the coordinate $x$ is located at a metastable state
($q=q_0+x$).
From this approximated potential,
the potential barrier height $V_0$ is calculated as
\begin{equation}
        V_0 = \sqrt{\frac{8\pi^3\eta^3}{\sqrt{3}\eps}}
                \simeq 12.0\eps^{-1/2} \eta^{3/2} .
        \label{V0value}
\end{equation}
The frequency of small oscillation around the metastable minimum
defined by $\omega_0=(V''(0)/m)^{1/2}$ is also obtained as
\begin{equation}
        \omega_0 = \left(\frac{2\pi\eps\eta}{3\sqrt{3}}\right)^{1/4}
                \simeq 1.05\eps^{1/4}\eta^{1/4} ,
        \label{w0value}
\end{equation}
where the fluxon mass $m=8$ has been substituted.
Note that $V_0$ and $\omega_0$ are normalized 
by $E_0$ and the plasma frequency $\omega_{\rm p}$, respectively.
For reference, we consider the exit point $x_0 (>0)$ defined by
$V(x_0)=0$. It is easily calculated as
\begin{equation}
        x_0 = \sqrt{\frac{9\sqrt{3}\eta}{8\pi\eps}}
                \simeq 0.79 \eta^{1/2}
                \eps^{-1/2}      . 
\end{equation}

The quantum decay rate from the metastable state at $T=0$
can be calculated by applying Langer's method to the effective action
(\ref{CaldeiraSeff}). The tunneling rate $\Gamma$ takes the form
\begin{equation}
        \Gamma = A\exp(-B) .
        \label{rate}
\end{equation}
The exponent $B$ is determined by the action of the nontrivial path 
$q_{{\rm B}}(\tau)$ which minimizes the effective action
(\ref{CaldeiraSeff}). This path is called a bounce solution.
As far as $x_0\ll1$, $q_{{\rm B}}(\tau)$ satisfies
\begin{equation}
        | q_{{\rm B}}(\tau) - q_{{\rm B}}(\tau') | \ll 1
\end{equation}
for all $\tau$, $\tau'$. Therefore, the assumption (\ref{condition}) 
is satisfied.  
The explicit forms of $A$ and $B$ are obtained only for the limiting cases.
In actual experiments, the damping coefficient 
$\alpha + \beta/3$ in eq.~(\ref{EOMF}) is small 
($\ll 0.1$ in the present situation). Then, the estimate in
the weak damping limit at $T=0$ is obtained by 
a perturbative treatment as~\cite{Caldeira}
\begin{eqnarray}
        A &=& \sqrt{60}\omega_0\omega_{\rm p}
        \left(\frac{B}{2\pi}\right)^{1/2} 
        (1 + {\cal O}(a)), 
        \label{prefactor} \\
        B &=& \frac{36V_0E_0}{5\hbar\omega_0\omega_{\rm p}} \left(
                1 + 1.74 a + {\cal O}(a^2) \right) \nonumber \\
        &\simeq& \frac{82.2\eta^{5/4}\eps^{-3/4}}{g^2}
        (1+1.74a+{\cal O}(a^2))
        \label{exponent} 
\end{eqnarray}
where $a=(\alpha+\beta/3)/2\omega_0$.
Since the predominant effect of the dissipation appears in the exponent
of the decay rate, the correction for the prefactor 
due to the dissipation is neglected in this paper.

Since $g^2$ takes a small value as given in $\S$~\ref{parameters}, 
the exponent $B$ takes a fairly large value.
To observe MQT of a fluxon in 
a time scale of the laboratory, 
we need to take a value of the parameter $\eta$
small to increase the decay rate.
For example, if we take $\eta = 5\times10^{-4}$, $\eps=0.1$,
then the values of $V_0$ and $\hbar\omega_0$ are estimated from
(\ref{V0value}) and (\ref{w0value}) as
\begin{eqnarray}
        V_0 E_0 &\sim& 0.82 W \, \mbox{[K]} \\
        \hbar\omega_0 \omega_{\rm p} &\sim& 0.34 \, \mbox{[K]} ,
\end{eqnarray}
in the original unit. 
We note that $\eta = 5\times10^{-4}$ and $\eps = 0.1$ are
not difficult to realize experimentally.
When $W=1[\mu\mbox{m}]$ and
dissipation effects are neglected
$(\alpha =\beta =0)$, the tunneling rate $\Gamma$ is
estimated from (\ref{rate}) with (\ref{prefactor}) and (\ref{exponent}) as
\begin{equation}
        \Gamma \sim 2 \times 10^4 [1/s] ,
\end{equation}
which is large enough to observe the MQT in the laboratory.
\subsection{Experimental design to observe MQT of a fluxon}
\label{apparatus}
\begin{figure}[tbp]
\hfil
\epsfile{file=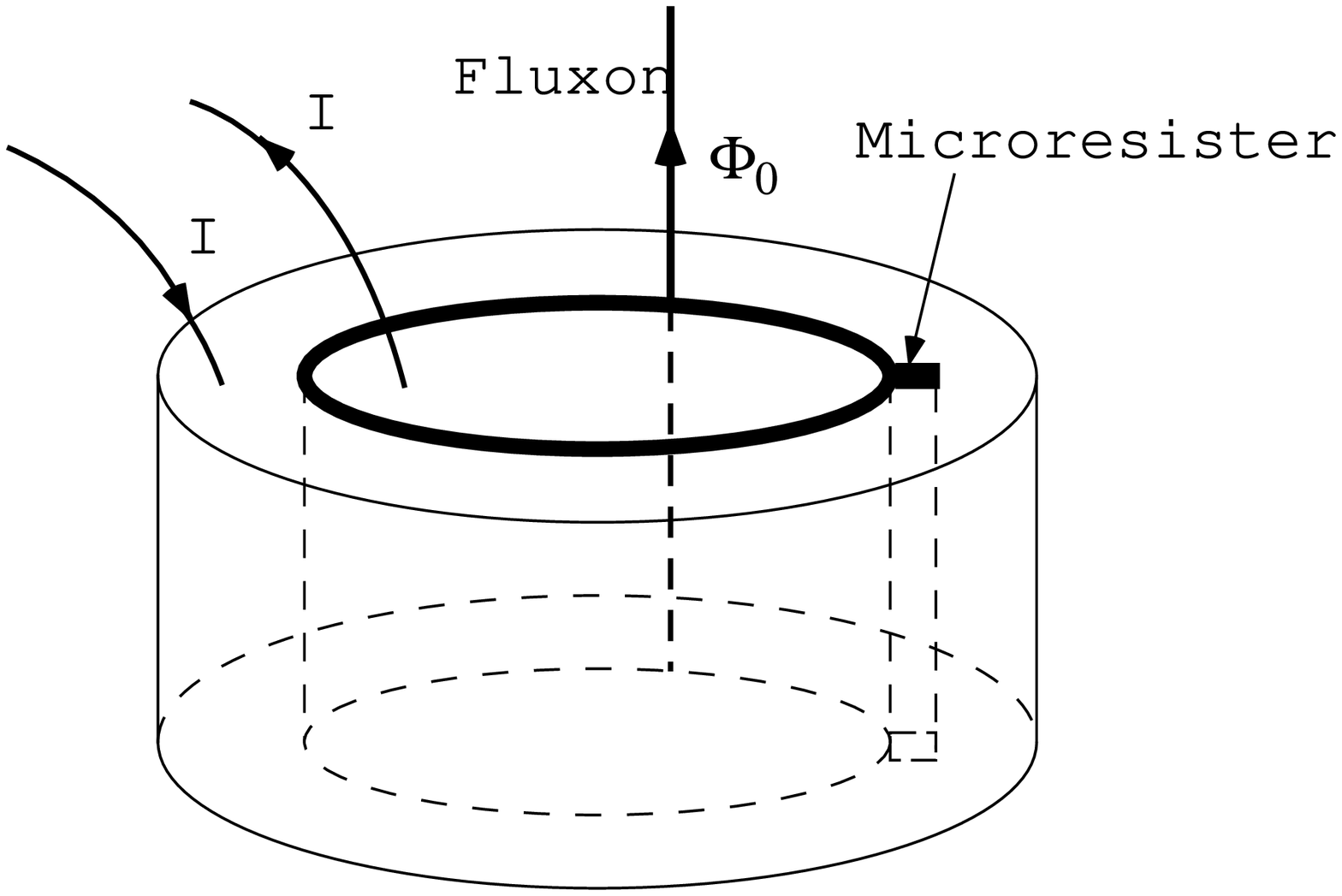,scale=0.6}
\hfil
\caption{A schematic configuration of the
proposed experimental apparatus. An external current $I$ is 
applied to the circular LJJ. Only one fluxon is captured through
the junction. A microresister is made in the junction.}
\label{expdrawing}
\end{figure}

In this subsection, we propose experimental apparatus to
observe the macroscopic tunneling of a single fluxon efficiently.
A schematic configuration of the considered Josephson junction is shown 
in Fig.~\ref{expdrawing}. 
Two superconducting cylinders
are separated by a thin insulating layer, where one fluxon is captured. 
The fluxon is accelerated by an externally driven current $I$, 
which is assumed to be spatially-uniform. A single microresister is made by
thickening the insulating layer locally. When $I=0$, the fluxon is
captured at the microresister. The critical current
of the junction is denoted by $I_0$.

Experiments on the junctions 
with a similar topological geometry have already been 
performed.~\cite{Davidson,Davidson2} 
In these experiments,
circular LJJs with the width $W\sim10[\mu\mbox{m}]$ have been
fabricated.
Further, by using scanning electron 
microscopy, it has been possible to introduce individual fluxons 
into such a system.~\cite{Ustinov}

\begin{figure}[tbp]
\hfil
\epsfile{file=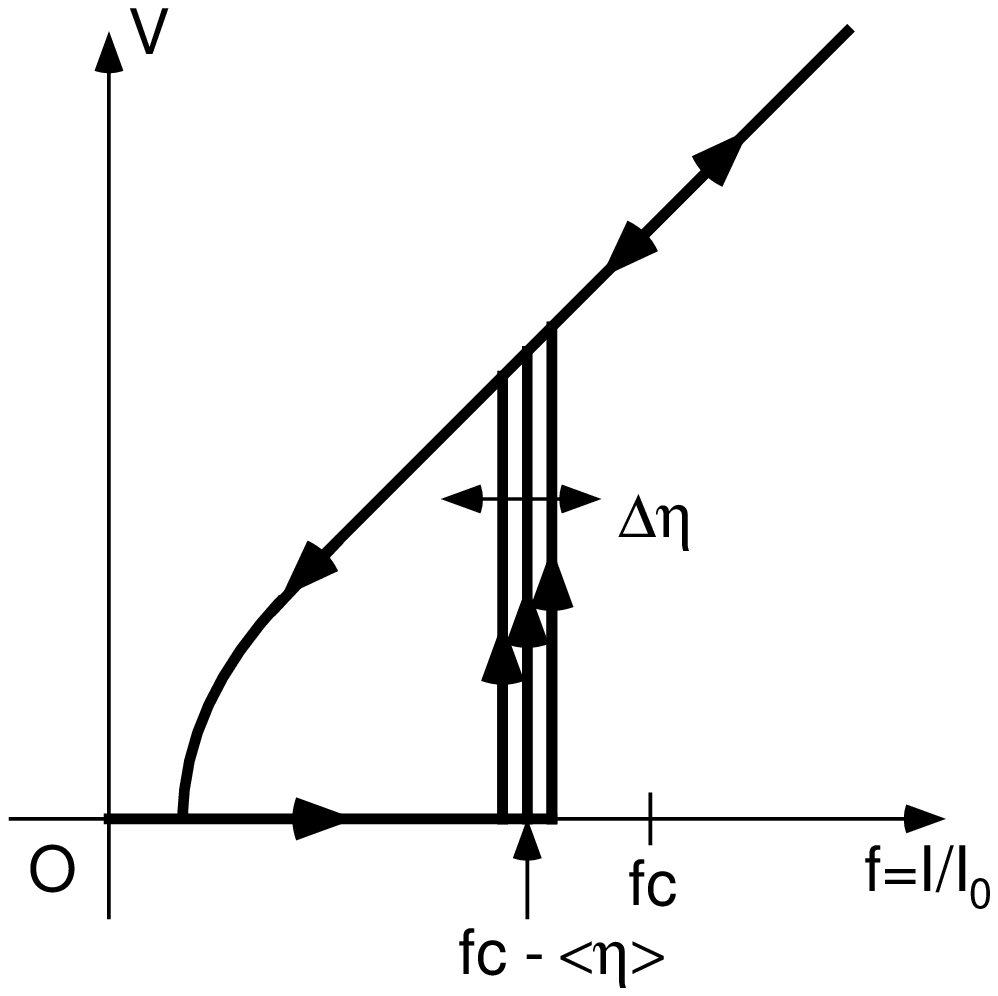,scale=0.8}
\hfil
\caption{The expected I-V curve of a circular LJJ. Depinning 
of a fluxon can be measured by switching to voltage states. 
The averaged switching current and the distribution width 
are denoted with $f-\langle \eta \rangle$ and $\Delta \eta$,
respectively. The tunneling rate $\Gamma(f)$ is obtained by the
form of the distribution.}
\label{IVcurve}
\end{figure}

The method to observe MQT of a fluxon 
may be very similar to the experiment 
technique used by Voss and Webb.~\cite{Voss}
When the external current $f =I/I_0$ is small, the fluxon stays
at a metastable state caused by the microresister.
When the current $f$ approaches the critical current 
$f_{\rm c}$ given by (\ref{fcdef}),
the metastable state vanishes and the fluxon becomes free to move. 
The depinning of a fluxon may occur before $f$ becomes $f_{\rm c}$ because of
the quantum tunneling.  After the depinning, moving fluxons generate
a voltage across the junction. Therefore, 
we can probe the depinning of a fluxon
by observing an I-V curve of the junction (See Fig.~\ref{IVcurve})
Since the tunneling is a stochastic process, the switching current
to voltage states has a distribution.
By applying an alternating current, 
a lot of fluxon-depinning events can be observed. 
(The moving fluxon is captured
in repeated intervals which satisfy $|f|\ll f_{\rm c}$.) 
The probability distribution of the depinning current $P(f)$ is related to
the tunneling rate $\Gamma(f)$ as 
\begin{equation}
        P(f) {\rm d}f 
        = \Gamma(f) \left( 1- \int_0^f P(f') {\rm d}f' \right) {\rm d}t .
\end{equation}
By solving this equation for $P(f)$, we obtain 
\begin{equation}
        P(f) = \left| \frac{{\rm d}f}{{\rm d}t} \right|^{-1}
                \Gamma(f) \exp \left( 
                - \left|\frac{{\rm d}f}{{\rm d}t}\right|^{-1} 
                \int_0^f \Gamma(f') {\rm d}f' \right) ,
        \label{pf}
\end{equation}
where $|{\rm d}f/{\rm d}t|$ is a sweep rate of the external current,
which must be small compared with $\omega_{\rm p}$.
Thus, we may obtain the tunneling rate $\Gamma(f)$ by measuring
$P(f)$ experimentally.~\cite{Voss}

The average and the mean-square value of the depinning current are expressed
by $f_{\rm c}-\langle \eta \rangle$ and $\Delta \eta \equiv 
\langle \eta^2 \rangle - \langle \eta \rangle^2$, respectively. 
The values of $\langle \eta \rangle$ and 
$\Delta \eta$ depend on the form of $\Gamma(f)$.
In this section, we only focus on $\Delta \eta$.
 
\subsection{Simulation of the distribution of depinning current}
To observe MQT of a fluxon in the experiments proposed 
in the previous subsection, 
several experimental requirements must be satisfied.
Among them, requirements for two physical quantities are important.
One is the current accuracy to measure $\Delta\eta$, 
and the other is the temperature $T$.

\begin{figure}[tbp]
\hfil
\epsfile{file=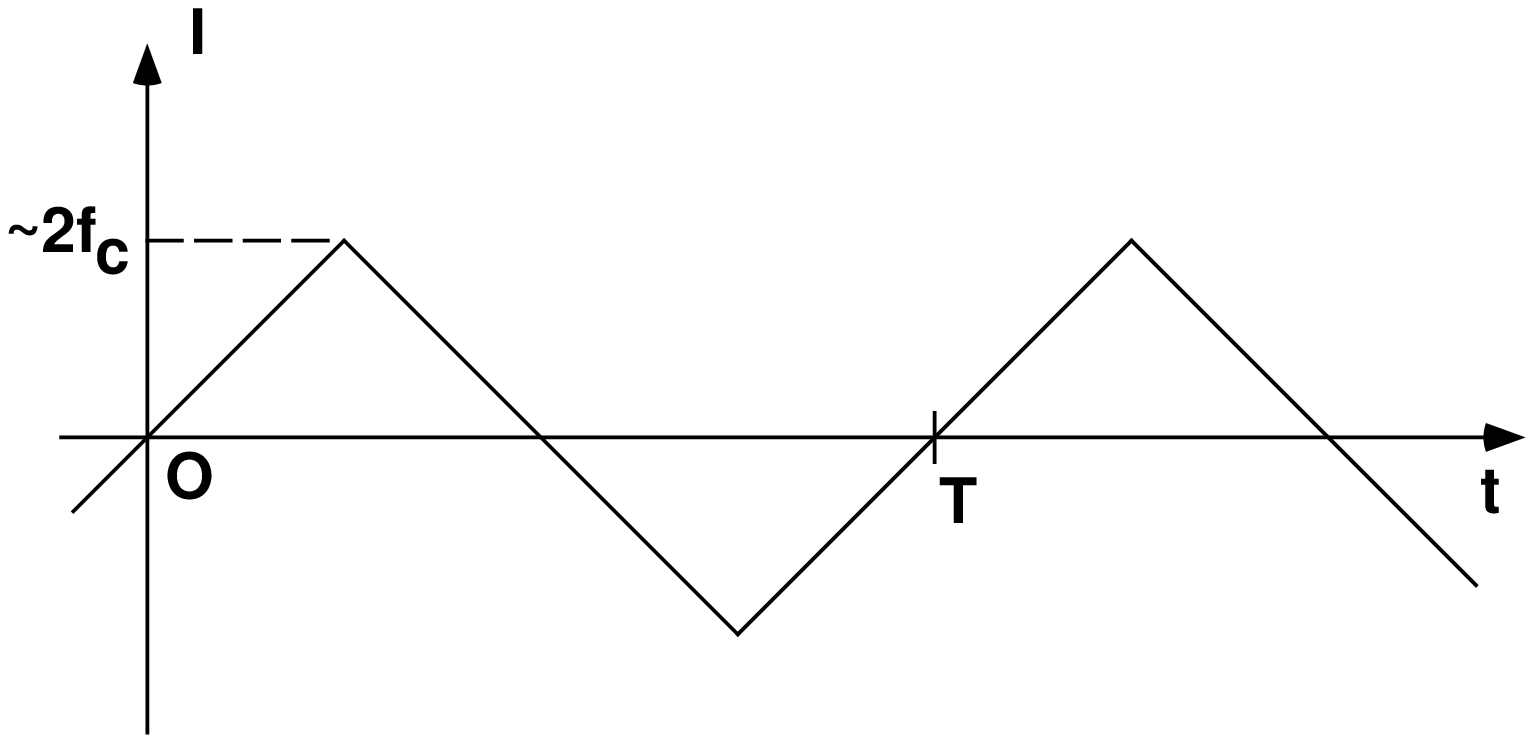,scale=0.6}
\hfil
\caption{The form of the external current used in the estimate.
The amplitude and the frequency of the alternating current is
taken as $2f_{\rm c}$ and $1/T = 15 \, [1/\mbox{s}]$, respectively.}
\label{currentform}
\end{figure}

To examine whether the observation of MQT is possible,
we first estimate
$\Delta \eta$ by using the estimated values given in $\S$~\ref{parameters}. 
We perform numerical integration of (\ref{pf}) to obtain $P(f)$. 
The tunneling rate $\Gamma(f)$ is obtained by substituting (\ref{g2value})
to (\ref{prefactor}) and (\ref{exponent}) in the original unit as 
\begin{eqnarray}
        \Gamma(f) &\simeq& 3.1 B^{1/2}
                \omega_0 \omega_{\rm p} \exp (-B) , 
        \label{rate-est} \\ 
        B &\sim& 4.1\times 10^4 \, \eta^{5/4} \eps^{-3/4} W 
        (1 + 1.74a) , \label{exponent-est}\\
        \omega_0 &\simeq& 1.05\eps^{1/4}\eta^{1/4},
\end{eqnarray}
where $\omega_{\rm p}$ is estimated in eq.~(\ref{wpdef}), and $a = (\alpha
+\beta/3)/2\omega_0$ is assumed to be small. 
Here $\eps$ is taken as 0.1 throughout this paper, 
which may be realized by thickening the insulator layer
locally with zero critical current density 
in a range of $0.1\lambda_{\rm J} \sim 2.7\,[\mu\mbox{m}]$,
which can be designed in the present available technique.
For convenience, we take a serrated wave as an alternating current form
as shown in Fig.~\ref{currentform}. 
The amplitude of the current
is taken as twice of $f_{\rm c}$ and the frequency is taken as
$1/T = 15 \, [1/\mbox{s}]$ to be compared with the experiment
performed by Voss and Webb.~\cite{Voss} In this case, we obtain
the sweep rate, $|{\rm d}f/{\rm d}t|= 120 f_{\rm c} \, [1/\mbox{s}]$.

Numerical calculation of (\ref{pf}) for $W=1 \, [\mu\mbox{m}]$
and $\eps = 0.1$ is performed for two cases;
we call the first a strongly dissipative 
case ($\alpha+\beta/3=0.091$), and the second 
a dissipationless case ($\alpha=\beta=0$).
Since the dissipation coefficient $\alpha + \beta/3$ becomes 
very small at low temperatures in actual experiments, 
we expect that the distribution of the depinning 
current is almost the same as dissipationless case.
To study dissipation effects, however, we also consider 
the strongly dissipative case, in which the damping coefficients
are taken as their upper bound estimated in $\S$~\ref{parameters}.
The value of $\Delta \eta$ in the strongly dissipative case 
gives the lower bound for $\Delta \eta$ in actual experiments.

\begin{figure}[btp]
\hfil
\epsfile{file=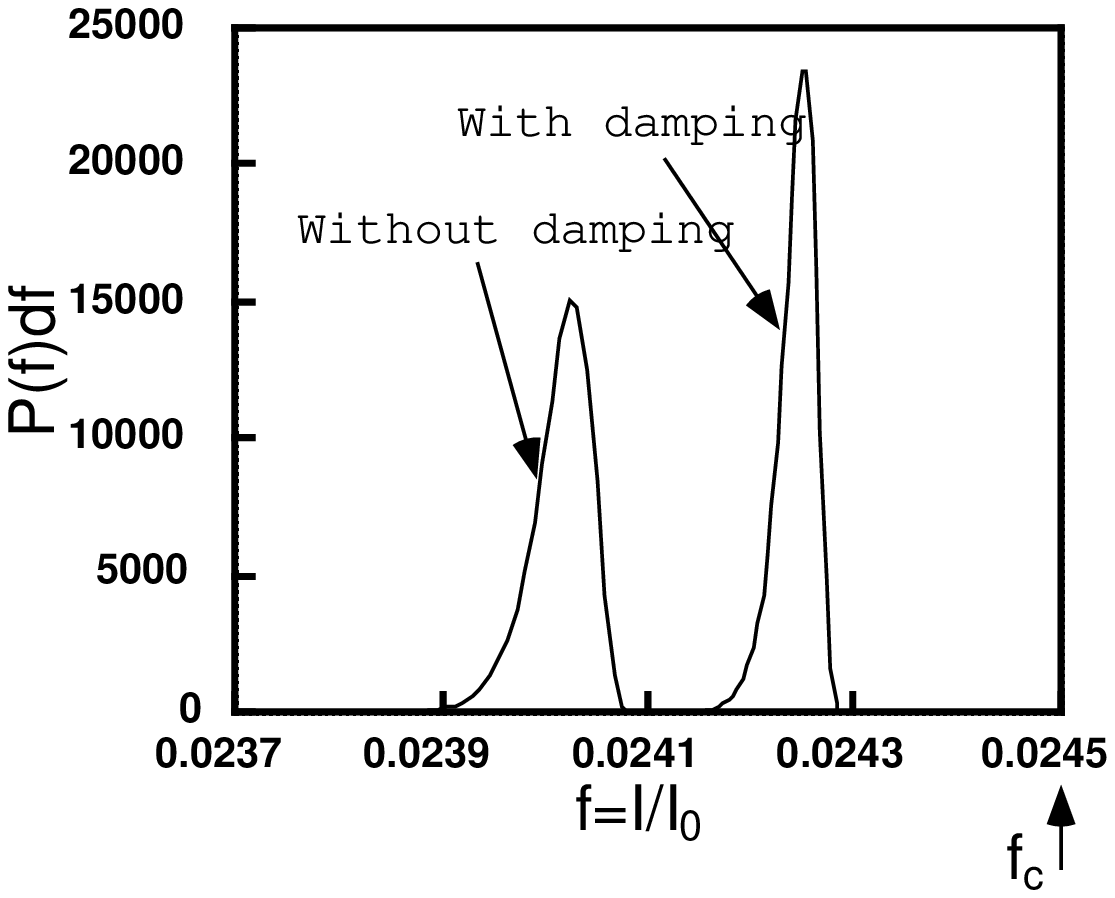,scale=0.8}
\hfil
\caption{The distribution function $P(I)$
with and without damping
obtained by numerical integration of (\ref{pf}) for $j_{\rm c}=3\times
10^6\,[\mbox{A}/\mbox{m}^2]$. The distribution widths are given by
$\Delta \eta = 2.0\times10^{-5}$, 
$\langle \eta \rangle = 2.6\times10^{-4}$
for the strongly dissipative case, and $\Delta \eta = 3.1\times10^{-5}$,
$\langle \eta \rangle = 4.9\times10^{-4}$ for the dissipationless case,
respectively. The classical depinning current $f_{\rm c}$ 
is 0.0245 in this estimate.}
\label{distribution}
\end{figure}
The obtained distribution is shown in Fig.~\ref{distribution}.
From the distribution, we obtain 
$\Delta \eta = 2.0\times10^{-5}$, 
$\langle \eta \rangle = 2.6\times10^{-4}$
for strongly dissipative case, 
and $\Delta \eta = 3.1\times10^{-5}$,
$\langle \eta \rangle = 4.9\times10^{-4}$ for dissipationless case. 
We find that $\Delta \eta$ is suppressed by dissipation effects.
We also find that dissipation effects on measurement of
the quantum tunneling are not small but modest.
For the strongly dissipative case, 
necessary accuracy of the current measurement 
is given by $\Delta \eta / f_{\rm c} \sim 8\times 10^{-4}$. 
On the other hand, in the experiment performed by Voss and 
Webb,~\cite{Voss} 
$\Delta I/I_0 < 2\times10^{-3}$ has already been realized. 
Here $\Delta I$ is the distribution width of
the switching currents to the voltage states, and $I_0$ is the
critical current. Therefore, if circular LJJs with 
$W=1[\mu\mbox{m}]$ can be fabricated,
accuracy of the current measurement 
to observe $\Delta \eta$ seems attainable
in the present available techniques.

\begin{figure}[tbp]
\hfil
\epsfile{file=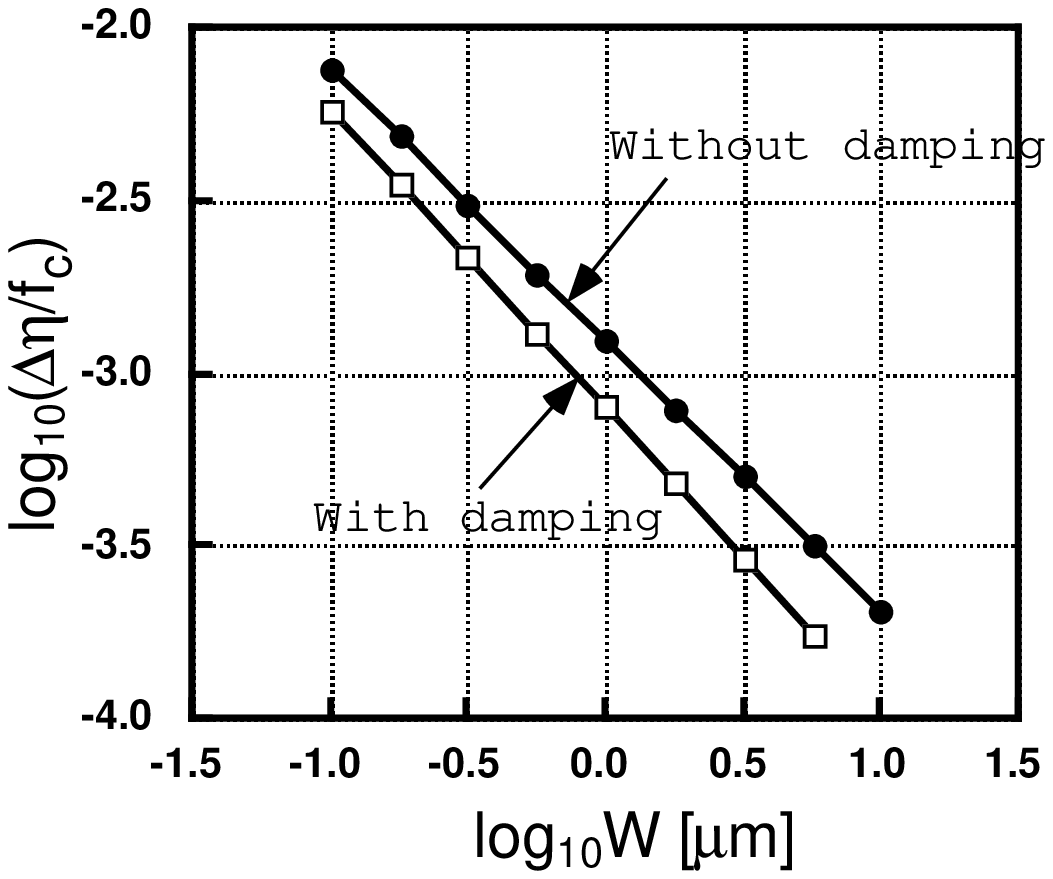,scale=0.8}
\hfil
\caption{The theoretical $W$-dependence of $\Delta \eta$ with and without
damping for $j_{\rm c} = 3\times10^6\,[\mbox{A}/\mbox{m}^2]$, 
$f_{\rm c} = 0.0245$. Fitting gives
$\Delta \eta \propto W^{-0.863}$ for the strongly dissipative case, and
$\Delta \eta \propto W^{-0.787}$ for the dissipationless case.}
\label{WIplot}
\end{figure}
If we change the junction width $W$, the distribution width 
$\Delta \eta$ takes a different value. The $W$-dependence of 
$\Delta \eta$ is shown in Fig.~\ref{WIplot} for
both strongly dissipative case and dissipationless case.  
From this dependence, we find that when the junction width $W$ is taken
larger, more accurate measurement for the current is needed. 
On the contrary, when $W$ is taken smaller, 
$\Delta \eta$ becomes large, and the observation
of MQT of a fluxon becomes easier. 
In practical experiments, we should take a proper value of $W$
so that the measurement of $\Delta \eta$ is possible.  
By fitting the estimated data, we obtain $\Delta \eta \propto W^{-0.787}$
for dissipationless case. The value of the power ($\sim -0.8$)
originates from the fact that the exponent is 
proportional to $\eta^{5/4}W$.
For strongly dissipative case, 
fitting gives $\Delta \eta \propto W^{-0.863}$.
As seen in Fig.~\ref{WIplot}, dissipation always suppresses
$\Delta \eta$, and the suppression becomes large
with the increase of $W$. This suppression is explained by the enhance
of the normalized damping strength $a=(\alpha+\beta/3)/2\omega_0$
due to the decrease of $\omega_0 \sim \langle \eta \rangle^{1/4}$.

Next, we consider finite temperature effects. At high temperatures,
thermally-activated decay occurs. The decay rate 
is formulated as
\begin{equation}
        \Gamma_{{\rm th}} = \frac{\omega_0}{2\pi} 
        \exp\biggl(-\frac{V_0}{k_{\rm B}T}\biggr) ,
        \label{rate-th}
\end{equation}
where $V_0$ is the energy barrier.
Here $V_0$ and the temperature $T$ are measured
in the original unit. 
It is expected that there exists a crossover temperature $T^*$  
which separates the thermal activated region and the quantum
tunneling region. Below $T^*$, the decay rate becomes 
independent of the temperature. The crossover temperature $T^*$
is defined as~\cite{Haenggi}
\begin{equation}
        T^* = \frac{\hbar\omega_{\rm b}}{2\pi k_{\rm B}}
        \biggl( ( 1 + \gamma^2)^{1/2} - \gamma \biggr) ,
\label{crt}
\end{equation}   
where $\gamma = \alpha + \beta/3$, and $\omega_{\rm b}$ is the frequency
at the top of the potential barrier. For the quadratic-plus-cubic potential,
$\omega_{\rm b}$ always agrees with $\omega_0$, which can be controlled
by the external current $f = f_{\rm c}-\eta$. We assume that
$T^*$ is averaged by the current sweep
to the value at $f=f_{\rm c}-\langle \eta \rangle$.
Then, we obtain the crossover temperature for $W=1\,[\mu\mbox{m}]$ 
as $T^*\sim54\,[{\rm mK}]$ for dissipationless cases, and
as $T^*\sim42\,[{\rm mK}]$ for strongly dissipative case.
Here, it appears that the suppression of $T^*$ due to damping is not
so large.

\begin{figure}[tbp]
\hfil
\epsfile{file=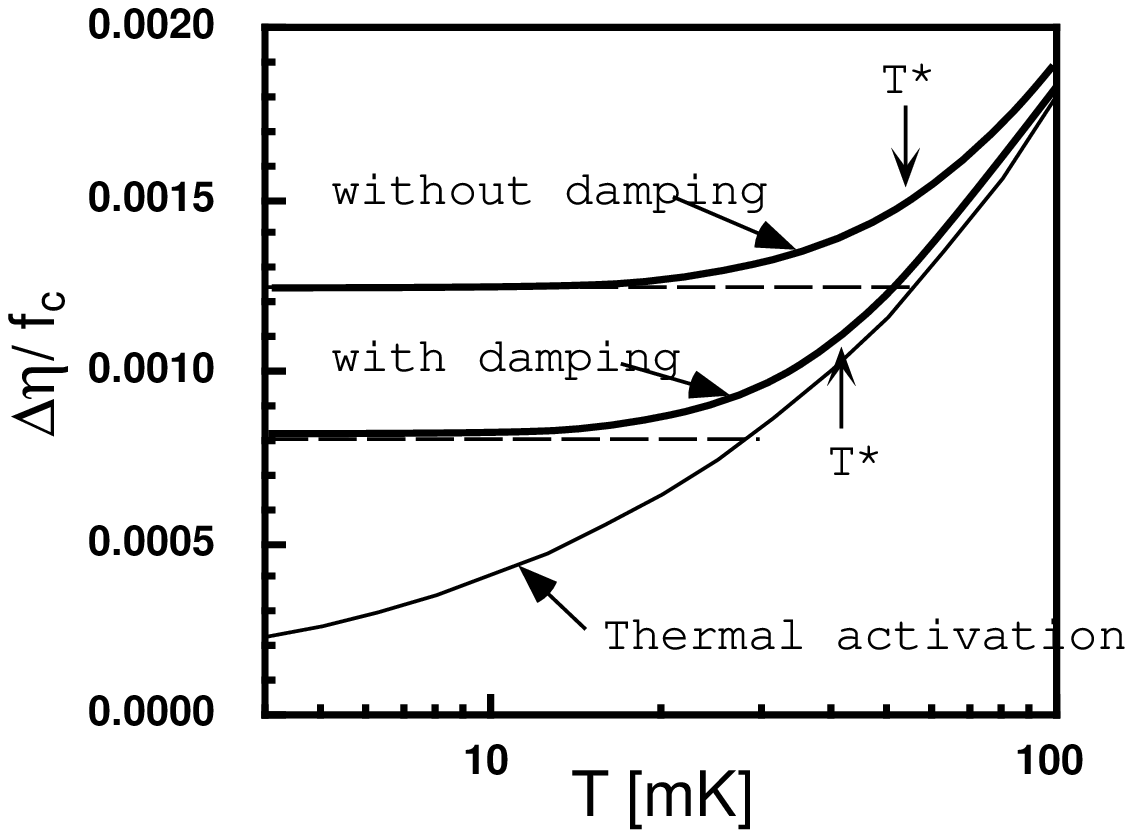,scale=0.8}
\hfil
\caption{The theoretical estimate for temperature-dependence 
of $\Delta \eta$ for the pure thermal activation and 
the pure MQT tunneling with and without damping. 
The behaviors of $\Delta \eta$ for the case with both effects 
are also shown by the bold thick curves. The crossover temperature
$T^*$ obtained by (\ref{crt}) is $42\,[\mbox{mK}]$ for the
strongly dissipative case, and $54\,[{\rm mK}]$ for the
dissipationless case.}
\label{TIplot}
\end{figure}

To observe MQT of a fluxon, the junction must be cooled
to low enough temperatures below $T^*$. 
We show in Fig.~\ref{TIplot}
theoretical widths $\Delta \eta$ expected from the pure 
thermal activation
by eq.~(\ref{rate-th}) as well as
from the pure MQT with and without damping
by eq.~(\ref{rate-est}) with (\ref{exponent-est}) 
for $W=1\,[\mu\mbox{m}]$.
The expected temperature-dependences of $\Delta\eta$
are also shown by the bold solid curve 
when both effects coexist.
From this calculation, 
it seems that the crossover temperature $T^*$ estimated by 
(\ref{crt}) gives a proper criterion.

\begin{figure}[tbp]
\hfil
\epsfile{file=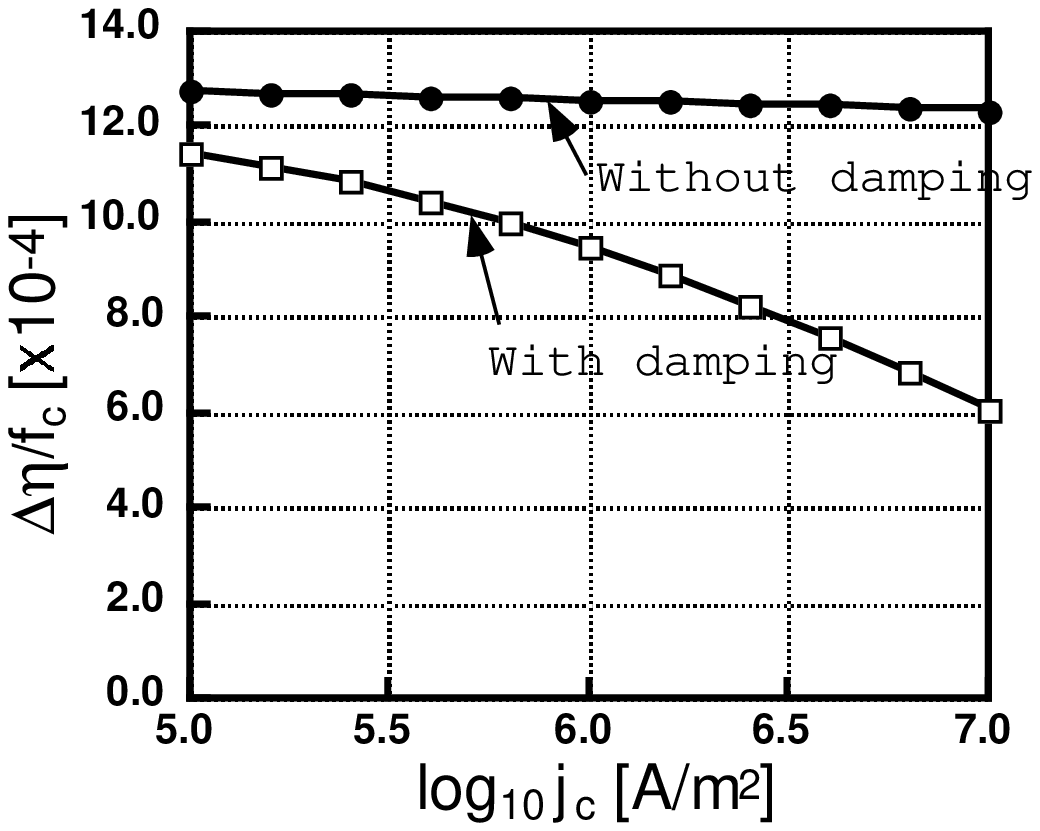,scale=0.8}
\hfil
\caption{The $j_{\rm c}$-dependence of $\Delta \eta$ with and without damping
for $W=1\,[\mu\mbox{m}]$.}
\label{JcIplot}
\end{figure}

Finally, we consider the case that the experimental parameters
given in Table~\ref{expparameter} are modified.
Since the accurate measurement of the parameters, $\lambda_L$,
$\eps_{\rm r}$, and $d$ is difficult, we only consider $j_{\rm c}$-control. 
The critical current density $j_{\rm c}$ can be changed
in a wide range by controlling the thickness of the oxide 
barrier.~\cite{foot1}
The $j_{\rm c}$-dependence of the estimated quantities is given 
from (\ref{jc1}), (\ref{wpdef}) and (\ref{jc2}) by
\begin{eqnarray}
        \lambda_{\rm J} &\propto& j_{\rm c}^{-1/2} , \\
        E_0 &\propto& j_{\rm c}^{1/2} , \\
        \omega_{\rm p} &\propto& j_{\rm c}^{1/2}.
        \label{wpdep}
\end{eqnarray}
Since $g^2 = \hbar \omega_{\rm p} / E_0$ is independent of $j_{\rm c}$,
we cannot control quantum fluctuations by $j_{\rm c}$.
Instead of it, the damping parameter $\alpha$
can be controlled by $j_{\rm c}$. When the gap of a superconductor $\Delta_0$ 
is fixed, $j_{\rm c}$-dependence of $\alpha$ is obtained from 
(\ref{wpdef}) and (\ref{aest}) as
\begin{equation}
        \alpha \propto j_{\rm c}^{1/2}.
        \label{ajc}
\end{equation}
According to this relation, the dissipation effects
get important with the increase of $j_{\rm c}$. 
The $j_{\rm c}$-dependence of $\Delta \eta$ for $W=1\,[\mu\mbox{m}]$ 
with and without damping is shown in Fig.~\ref{JcIplot}.
The widths $\Delta \eta$ are almost independent of $j_{\rm c}$
in the dissipationless case, while $\Delta \eta$
is suppressed in the strongly 
dissipative case with the increase of $j_{\rm c}$.

\begin{figure}[tbp]
\hfil
\epsfile{file=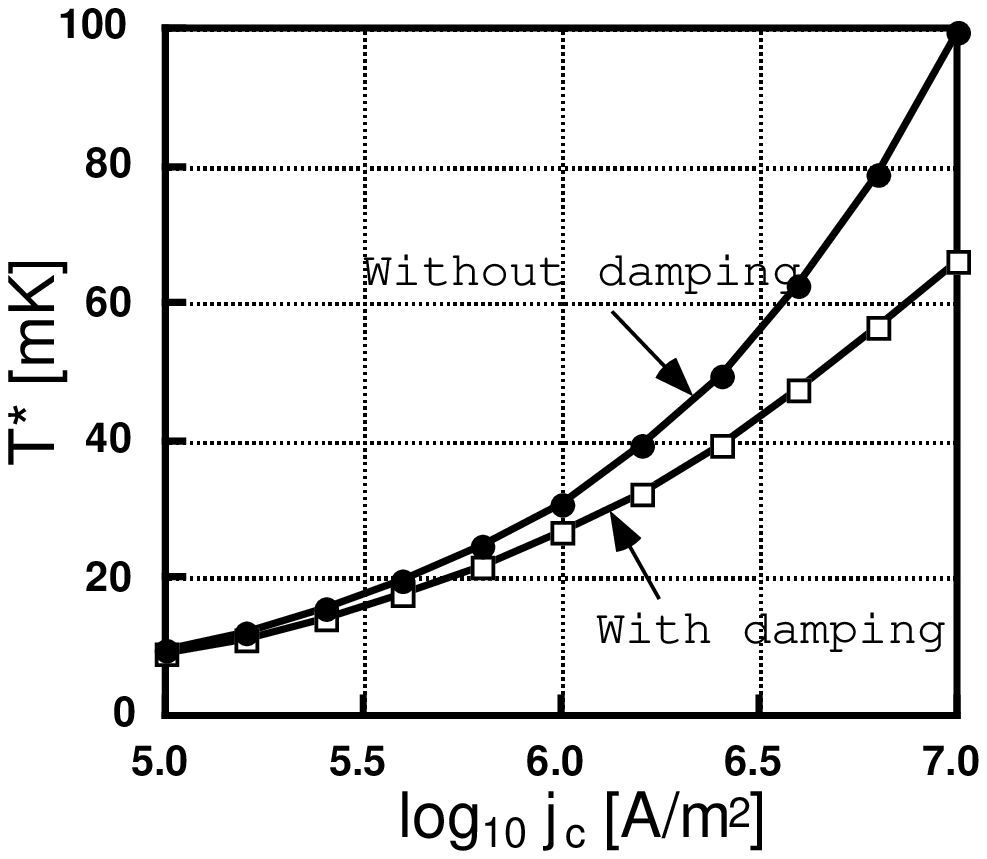,scale=0.8}
\hfil
\caption{The $j_{\rm c}$-dependence of the crossover temperature $T^*$ 
with and without damping for $W=1\,[\mu\mbox{m}]$.}
\label{C-temp}
\end{figure}

From (\ref{wpdep}), the crossover temperature 
$T^* \propto \omega_{\rm p}$ is also controlled
by $j_{\rm c}$. The $j_{\rm c}$-dependence of $T^*$ calculated from (\ref{crt})
with and without damping for $W=1\,[\mu\mbox{m}]$ 
is shown in Fig.~\ref{C-temp}.
Fitting the data without damping gives $T^* \propto j_{\rm c}^{0.506}$,
which is nearly the value expected from (\ref{wpdep}).
The deviation originates from the small change in $\langle\eta\rangle$.
In the strongly dissipative case, 
$T^*$ is suppressed, and the suppression
from the dissipationless case
grows with the increase of $j_{\rm c}$, because the damping effects 
become important through (\ref{ajc}). 
At any rate, however, the absolute value of the
crossover temperature $T^*$ is enhanced 
with the increase of $j_{\rm c}$. Hence, $j_{\rm c}$ should be taken large 
to observe MQT of a fluxon more easily
in the range of $j_{\rm c}$ given in Fig.~\ref{C-temp}.
\section{Tunneling in a Two-State System}
\label{twostate}
In this section, we study a two-state system made by two microresisters. 
In classical mechanics, a fluxon may stay at either of two stable
states at zero temperature. In quantum mechanics, however, 
quantum tunneling through the energy barrier is possible. 
When the energy levels of
the ground state at each well is the same and dissipation is neglected,
the quantum tunneling makes an energy splitting and generates 
oscillation of a fluxon between the two wells retaining the coherence. 
This macroscopic effects is called quantum macroscopic coherence (MQC).

In the last ten years, MQC has been investigated theoretically
in detail.~\cite{Leggett1} MQC is the key phenomenon
to clarify the validity of quantum mechanics at macroscopic 
level.~\cite{Leggett2} 
MQC is far more sensitive to environmental suppression
than MQT because a coupling between the relevant macroscopic 
variable and its environment rapidly destroys the phase
coherence between two states, while we need to retain phase coherence
for much longer time for MQC than MQT. 
Thus MQC is in general much harder to observe.
Several experiments have been proposed to observe MQC 
on current biased Josephson junctions and SQUIDs. 
However, to date, there is no evidence for MQC in spite of
substantial effort.
Here, we mainly focus on dissipation effects on MQC in LJJs, 
and pursue possibility to observe MQC in this system. 

When there exists no external current ($f=0$), 
we obtain the potential made by two microresisters as
\begin{equation}
        V(q) = -\frac{2\eps}{\cosh^2(q-l/2)}
               -\frac{2\eps}{\cosh^2(q+l/2)} ,
\end{equation}
where the origin of $q$ is taken at the midpoint of the resistors, and
the distance between the resistors is denoted with $l$.
The potential $V(q)$ has only one stable state for small $l$,
while the potential has a double well structure, when $l > l_0$.
The critical length $l_0$ is given by
\begin{equation}
        l_0 = \ln \left(\frac{\sqrt{3}+1}{\sqrt{3}-1}\right) 
        \simeq 1.317 .
\end{equation}
We only consider the case that the potential barrier is small. 
In this situation, we may assume $l=l_0+a$ with $a\ll1$.
The potential term can be expanded around $q=0$ as
\begin{equation}
        V(q) = \frac{32}{27} \eps q^4 - \frac{16}{3\sqrt{3}}\eps a q^2.
        \label{approxpot22}
\end{equation}
From this potential, the position of the stable states is obtained
as $q=\pm q_0$ where
\begin{equation}
        q_0 = \sqrt{\frac{3\sqrt{3}}{4} a} \simeq 1.14a^{1/2} .
\end{equation}
When $a \ll 1$, we obtain $q_0 \ll 1$ and the expansion
in the form of the potential (\ref{approxpot22}) is valid around the two
stable states.

The amplitude of the dissipation is 
determined by a dimensionless quantity $K$ as
\begin{equation}
        K = m \biggl(\alpha + \frac{\beta}{3} \biggr)
        \frac{(2q_0)^2}{2\pi g^2}.
\end{equation}
Here $m=8$ is the mass of a fluxon.
By using the estimated values of $g^2$, 
we estimate $K$ as
\begin{equation}
        K \sim 3.3\times10^3 
        \biggl( \alpha + \frac{\beta}{3} \biggr)a W .
\end{equation} 
Note that $K$ is independent of details of the potential form
and determined by the distance between potential minima $2q_0$ 
and the width of the junction $W$.

To estimate $K$, we assume that the ratio between $\alpha$ and $\beta$ 
is independent of temperatures.~\cite{Davidson,Davidson2} 
Then, because $\beta$ is not dominant in the dissipation 
amplitude $\alpha + \beta/3$ at relatively high
temperatures as is observed experimentally,~\cite{Pedersen2} 
we expect that $\beta$ is not
dominant also at low temperatures.
Hence we neglect the $\beta$-term, and estimate the value 
of $K$ roughly as 
\begin{equation}
K \sim 3\times 10^2 aW\frac{\alpha}{\alpha_{\rm n}} 
= 3\times 10^2 aW\frac{r_{\rm n}}{r_{\rm qp}} .
\end{equation}
Here, we inserted the dissipation coefficient 
$\alpha_{\rm n} = 0.088$ estimated in $\S$~\ref{parameters}.
The ratio between the normal resistance $r_{\rm n}$ and
the quasiparticle resistance $r_{\rm qp}$ can be obtained
experimentally.

The condition to observe MQC is given as~\cite{Leggett1,Leggett2}
\begin{equation}
        K \ll 1 \quad \mbox{and}
        \quad k_{\rm B} T \ll \hbar\Gamma / K.
        \label{condMQC}
\end{equation}
If $a=0.1$, $W=1$, and $r_{\rm n}/r_{\rm qp}=10^{-3}$ 
can be realized, the estimated value of $K (\sim 0.03)$
seems to satisfy the first conditions in (\ref{condMQC}). 
Although the estimate of the ratio $r_{\rm n}/r_{\rm qp} = 10^{-3}$ 
is very rough, it seems to be attainable at low temperatures 
of mK order.~\cite{Pedersen2}
Hence, we expect that the observation of MQC in LJJs appears to
be possible within the Caldeira Leggett theory.


\section{Concluding Remarks}
\label{discussion}
In this paper, we have studied two kinds of quantum tunneling
for a single fluxon. We have
found that the quantum tunneling from a metastable state
may be observed at low temperatures (mK order), and that
the junction width should be as small as possible.
If the junction width is taken as $\sim 1 [\mu\mbox{m}]$, 
required accuracy of current measurement
seems attainable in the laboratory. 
We have also found that the observation of MQC in a two-state system
in LJJs appears to be possible, because dissipation
due to the quasiparticle tunneling is strongly suppressed
at low temperatures. It should be noted, however, that the damping
amplitude may increase by other dissipation sources.
In the experiment on SQUIDs,~\cite{Han} the observed damping
amplitude is much larger than one estimed by quasiparticle resistance.
In addition to MQC, we expect other macroscopic quantum effects
characteristic in a two-state system. For example, it may be possible
to observe the incoherent tunneling and 
the population inversion between quantum states in LJJs
as observed in SQUIDs.~\cite{Han,Han2}

The most characteristic feature of LJJs is that the phase difference
$\phi$ has a spatial dependence. In other words, $\phi$ is
a field variable with infinite degrees of freedom.
This opens the possibility of studying combined effects of
macroscopic quantum phenomena and many-body effects beyond
the phenomenological theory by Caldeira and Leggett.
In this paper, however, dynamics of only one degree of freedom, i.e.,
the center position of the fluxon has been considered
as the first attempt.
Other infinite degrees of freedom of the field appears
in the form of plasmons. They, however, do not play an important role
in quantum dynamics of a fluxon because of two reasons: (i)
plasmon excitation has an energy gap,
and plasmons are suppressed at low temperatures;
(ii) in the lowest order of $g^2$, plasmons are decoupled
from the fluxon. Hence, within the present study, 
characteristic features of the field with infinite degrees of freedom
do not appear. To study many-body effects of the field variable,
we must consider different situations.
Fortunately, it is rather easy to devise LJJs compared with other systems
described by field theories. New and rich quantum phenomena 
may appear by such devices, and it will be a challenging subject
to study the interplay of many-body interaction 
and the macroscopic quantum effects.

\acknowledgement

We would like to thank A. Shnirman for pointing out 
strong suppression of dissipation coefficients at low temperatures.

\appendix
\section{Derivation of the equation of motion} \label{sfapp1}
In this appendix, we derive the classical equation of motion (\ref{SGEI})
from the model Hamiltonian (\ref{fieldHtot}) with (\ref{Hs})-(\ref{Hb})
within the classical mechanics. First, we write the Hamilton equations,
\begin{eqnarray}
        \dot{\Pi}&=& -\frac{\delta H}{\delta \phi(x)} ,
        \label{Hameq1}  \\ 
        \dot p_j &=& -\frac{\delta H}{\delta q_j(x)} , 
        \label{Hameq2}  \\
        \dot p_j'&=& -\frac{\delta H}{\delta q_j'(x)} .
        \label{Hameq3}
\end{eqnarray}
Here $\Pi = \dot \phi$ is a momentum conjugate to $\phi$.
By substituting the Hamiltonians (\ref{Hs})-(\ref{Hb})
into these equations, we obtain series of equations,
\begin{eqnarray}
        & & \phi_{tt}-\phi_{xx} + \sin \phi + f 
        - \eps \delta(x) \sin \phi 
        + \frac{\delta H_{\alpha}}{\delta \phi(x)} 
        + \frac{\delta H_{\beta}}{\delta \phi(x)} = 0, 
        \\
        & & m_j\ddot q_j + m_j\omega_j^2q_j - c_j\phi = 0 ,
        \label{transs} \\
        & & m_j'\ddot q_j'+m_j'\omega_j'^2q_j' -c_j'\phi_x = 0 .
\end{eqnarray}
where
\begin{eqnarray}
        \frac{\delta H_{\alpha}}{\delta \phi(x)}
        &=& - \sum_j c_j \left( 
        q_j - \frac{c_j^2}{m_j \omega_j^2} \phi\right) , \\
        \frac{\delta H_{\beta}}{\delta \phi(x)}
        &=& \sum_j c_j' \left( 
        q_{j,x}'-\frac{c_j'^2}{m_j'\omega_j'^2} \phi_{xx}\right) .
        \label{sign}
\end{eqnarray}
Note that the sign is changed in (\ref{sign}) from 
the integration by part.
The Fourier transformations of $q_j$, $q_j'$ and $\phi$ 
are denoted by
\begin{eqnarray}
        \phi(x,t)&=&\int \frac{{\rm d}\omega}{2\pi} 
        \int \frac{{\rm d}k}{2\pi}
                \tilde{\phi}(k,\omega) 
                \mbox{e}^{\mbox{i}kx-\mbox{i}\omega t} , \\
        q_j(x,t)&=&\int \frac{{\rm d}\omega}{2\pi} 
        \int \frac{{\rm d}k}{2\pi}
                \tilde q_j(k,\omega) 
                \mbox{e}^{\mbox{i}kx-\mbox{i}\omega t} , \\
        q_j'(x,t)&=&\int \frac{{\rm d}\omega}{2\pi} 
        \int \frac{{\rm d}k}{2\pi}
                \tilde q_j'(k,\omega) 
                \mbox{e}^{\mbox{i}kx-\mbox{i}\omega t} .
\end{eqnarray}
Fourier transforms of (\ref{transs})-(\ref{sign}) 
and elimination of $\tilde q_j$ and $\tilde q_j'$ give
\begin{eqnarray}
        \widetilde{\frac{\delta H_{\alpha}}{\delta \phi}}
                &=& -\mbox{i}\omega
                \tilde{\alpha}(\omega)\tilde{\phi}(k,\omega) ,
        \label{aterm}\\
        \widetilde{\frac{\delta H_{\beta}}{\delta \phi}}
                &=& -\mbox{i}k^2\omega
                \tilde{\beta}(\omega)\tilde{\phi}(k,\omega) ,
        \label{bterm}
\end{eqnarray}
where
\begin{eqnarray}
\tilde{\alpha}(\omega) &=& -\frac{2\mbox{i}\omega}{\pi}
        \int_0^{\infty} \! \! {\rm d}\omega' 
        \frac{J_{\alpha}(\omega')}
        {\omega'(\omega'^2-\omega^2-\mbox{i}0^+)} ,\\
\tilde{\beta}(\omega) &=& -\frac{2\mbox{i}\omega}{\pi}
        \int_0^{\infty} \! \! {\rm d}\omega' 
        \frac{J_{\beta}(\omega')}
        {\omega'(\omega'^2-\omega^2-\mbox{i}0^+)} .
\end{eqnarray}
Here, the spectral functions $J_{\alpha}(\omega)$ and $J_{\beta}(\omega)$
are defined by (\ref{ja})-(\ref{jb}). The positions of poles 
are taken off the real axis to satisfy the causality.
When we choose the spectral functions as 
$J_{\alpha}(\omega) = \alpha \omega$, 
$J_{\beta}(\omega) = \beta \omega$, we find
that $\tilde{\alpha}(\omega)$ and 
$\tilde{\beta}(\omega)$ acquire real parts as 
$\tilde{\alpha}'(\omega)= \alpha$, 
$\tilde{\beta}'(\omega)=\beta$.
Finally the inverse Fourier transforms of (\ref{aterm}) and
(\ref{bterm}) give dissipative terms in the classical perturbed
sine-Gordon equation (\ref{SGEI}). 

\section{Details of the Semiclassical Theory}\label{sfapp2}
In this appendix, the semiclassical calculation 
is presented in detail. Here, we adopt a path integral method used by Gervais 
{\it et al.}~\cite{Gervais}
The functional integral method obscures problems of quantum 
operator ordering.~\cite{Tomboulis} Actually in the correct formalism
based on operator canonical transformations,~\cite{Chist}
additional terms, which is absent in the straightforward functional 
approach, appear in the expansion.
The extra terms are, however, irrelevant to the present
calculation because the correction appears only in higher order of $g^2$.

First we consider the unperturbed sine-Gordon equation given 
by (\ref{Z0defapp}) and (\ref{S0defapp}).
We introduce a new field variable $\vphi=\phi/g$ 
to perform the expansion of $g^2$ easily. 
The partition function can be modified as
\begin{eqnarray}
        & & Z_0 = \oint {\cal D} \vphi(x,\tau) {\cal D} \Pi(x,\tau)
        \exp\biggl[ \mbox{i}
        \int {\rm d}\tau \int {\rm d}x \Pi 
        \dot{\vphi} \biggr. \nonumber \\
        &-& \biggl. \int {\rm d}\tau \int {\rm d}x 
        \Bigl( \frac12 \Pi^2 +\frac12 \vphi_x^2 
        + \frac{1}{g^2} (1-\cos g\vphi) \Bigr) \biggr]  ,
\end{eqnarray}
where $\Pi$ is a field momentum variable conjugate to $\vphi$. 

The stationary nontrivial solutions takes the form
\begin{equation}
        \vphi_0(x-q) = \frac{4}{g} \arctan \Bigl[ \exp (x-q) \, \Bigr] ,
        \label{SolitonSolution2}
\end{equation}
where $q$ is the center coordinate of the soliton. 
Then, we describe the field $\vphi$ as follows:
\begin{equation}
        \vphi(x,\tau) = \vphi_0(x-q(\tau)) + \eta(x-q(\tau),\tau) .
\end{equation}
Since $\vphi$ is a quantum field, it has an infinite number 
of degrees of freedom in the functional integral.
Thus, $\eta(x-q(\tau),\tau)$ also
has an infinite number of quantum degrees of freedom, and
$\vphi_0(x-q(\tau))$ contains only one quantum 
degree of freedom in $q(\tau)$.
In order to keep the number of degrees of freedom, we should
set a subsidiary condition for $\eta$,
\begin{equation}
        \int {\rm d}\rho \vphi_0'(\rho) \eta(\rho,\tau) = 0 .
\end{equation}
Here $\rho = x - q(\tau)$ is a soliton-fixed coordinate,
and the prime denotes derivative with respect to $x$.
In a proper way, we can define momentum variables
$p(\tau)$, $\pi(\rho,\tau)$ conjugate to $q(\tau)$, 
$\eta(\rho,\tau)$. The partition function
is modified as~\cite{Gervais}
\begin{full}
\begin{eqnarray}
        Z &=& \oint {\cal D}p(\tau) {\cal D}q(\tau)
                {\cal D}\pi(\rho,\tau) {\cal D}\eta(\rho,\tau)
                \delta(\int {\rm d}\rho \: \vphi_0'\eta)
                \delta(\int {\rm d}\rho \: \vphi_0'\pi ) 
        \nonumber \\
        & & \hspace{5mm}
                \exp\left(\mbox{i}\int {\rm d}\tau \: p\dot{q} 
                + \mbox{i} \int d\tau \int {\rm d}\rho \: \pi \dot{\eta} 
                - \int {\rm d}\tau H \right) ,
\end{eqnarray}
where
\begin{eqnarray}
        H &=& M_0 + \frac{(p+\int {\rm d}\rho \: \pi \eta')^2}
                {2M_0(1+\xi/M_0)^2} 
                +\int {\rm d}\rho \: \frac12 \left(
                        \pi^2+\eta'^2+(1-\frac{2}{\cosh^2 \rho})\eta^2
        \right) \nonumber \\
        &-& \int {\rm d}\rho \left(
                \cos g(\vphi_0 +\eta) - \cos g\vphi_0
                + g\eta \sin g\vphi_0
                + \frac12 g^2 \eta^2 \cos g\vphi_0 \right) . 
\end{eqnarray}
\end{full}
Here $M_0 = 8/g^2$, and
\begin{equation}
        \xi = \int {\rm d}\rho \vphi_0'(\rho) \eta'(\rho,\tau).
\end{equation}
We divide the Hamiltonian into three parts,
\begin{eqnarray}
        H_1 &=& M_0 + \frac{p^2}{2M_0} , \\
        H_2 &=& \int {\rm d}\rho \: \frac12 \left(\pi^2 +\eta'^2 
                +(1-\frac{2}{\cosh^2 \rho}) \eta^2 \right) , \\
        H_3 &=& H-H_1-H_2 .
\end{eqnarray}
Here $H_1$ describes fluxon dynamics in the lowest approximation.
The quadratic part $H_2$ is easily quantized, and gives
plasmon excitations. These excitations have a gap of $\hbar \omega_{\rm p}$
in the original unit. 
Taking $H_1+H_2$ as the unperturbed part, we can perform
the traditional perturbation expansion for $H_3$,
which includes only higher order terms than quadratic.
This expansion for $g^2 \ll 1$ is obtained by
Gervais {\it et al.}~\cite{Gervais} in detail.
The result is
\begin{eqnarray}
        & & Z_0 = Z_{\rm p} \oint {\cal D}q(\tau){\cal D}p(\tau)\exp\left[
                \mbox{i}\int {\rm d}\tau p(\tau) \dot q(\tau) \right.
        \nonumber \\
        & & - \left. \int {\rm d}\tau \biggl( \frac{m}{g^2} 
                + \frac{g^2p(\tau)^2}{2m} + \Delta E[p(\tau)] \biggr)
                \right] , 
        \label{semiclmethod}
\end{eqnarray}
where $m$ is the soliton mass which equals identically to 8, and
$Z_{\rm p}$ is the partition function of unperturbed plasmons,
which is irrelevant to the tunneling rate.
The quantum correction $\Delta E$
can be calculated in a systematic way
using the traditional perturbation theory represented 
by familiar graphical diagrams. As shown in eq.~(\ref{semiclmethod}),
the lowest order approximation for $g^2$ is nonrelativistic.
The Lorentz-invariant form is obtained by summing up all relevant 
diagrams.~\cite{Gervais} The result is $\Delta E \sim {\cal O}(g^2)$,
and there exists no correction to the term proportional to $p^2g^2$
in $\Delta E$. 

When $g^2$ is small, $\Delta E$ can be neglected. 
After integrating (\ref{semiclmethod}) over $p$, we obtain
\begin{equation}
        Z_0 = \oint {\cal D}q(\tau) 
        \exp\left(-\frac{1}{g^2}\int {\rm d}\tau
        \Bigl( m +\frac m2 \dot q(\tau)^2 \Bigr) \right) .
\end{equation}



\begin{thebibliography}{99}
\bibitem{foot1}In addition to the $j_{\rm c}$-control by the 
thickness of oxide barrier, $j_{\rm c}$ can be controlled 
by the external magnetic fields
because of the change of the gap $\Delta_0$.
\end{thebibliography}
\end{document}